%% file: Submission22.tex
\RequirePackage[l2tabu,orthodox]{nag}
\documentclass[envcountsect,envcountsame]{eptcs}
\providecommand{\event}{QPL 2017}
\providecommand{\authorrunning}{Dunne}
\providecommand{\titlerunning}{On the Structure of Abstract 
$H^*$--Algebras}
\usepackage{mathrsfs} 
\usepackage{amsopn}

\input{preamble.tex}
\usepackage[undirected]{tikzquanto}
\begin{document}
\input{front-matter.tex}

\maketitle

\input{abstract.tex}

\section{Introduction}
\label{sec:Introduction}

The 
present work is part of an 
ongoing project \cite{Dunne2016:NewPerspective,Dunne2017:SpecPreshKSAndQVR} to 
bridge the monoidal approach to quantum 
theory of Abramsky and Coecke \cite{AbramskyCoecke2004:CategoricalSemantics}, 
and the topos approach 
to quantum theory of Butterfield, Doering and Isham
\cite{IshamButterfield1998:AToposPerspective,DoeringIsham2008:WhatIsAthing}. 
Both the 
monoidal and topos approaches to 
quantum theory are algebraic, in that they seek to represent some aspect of 
physical reality with algebraic structures. By taking the concept of a 
``physical observable'' as a fixed point of reference we cast the difference 
between 
these approaches
as internal vs. external 
points of view. In particular, in the monoidal approach one encodes the notion 
of ``observable'' as an internal commutative algebra (for example a Frobenius 
algebra \cite{CoeckeEtAl2008:NewDescriptionOrthogonal} or an $H^*$--algebra 
\cite{AbramskyHeunen2012:HAlgebras}) in some suitable monoidal 
category $\catA$ -- traditionally the category $\Hilb$ of Hilbert spaces and 
bounded linear maps. The topos approach to quantum theory, as presented in 
\cite{Flori2013:Topos}, considers 
representations of commutative algebraic structures (for example 
$C^*$--algebras, or von Neumann algebras 
\cite{Conway2000:ACourseInOperatorTheory}) on $\Hilb$. The topos approach makes 
essential use of the fact that the sets $\Hom(H,H)$ for $\Hilb$ carry the 
structure of a $C^*$--algebra. In 
\cite{Dunne2016:NewPerspective} we 
showed that the categories considered in the monoidal approach have a similarly 
rich algebraic structure on their sets of endomorphisms $\Hom(A,A)$, allowing 
one to take the external perspective for a broad class of categories, not just 
$\Hilb$. Here we show that there is a natural way 
to pass from the internal algebraic structures which represent observables to 
external algebraic representations of observables, and hence we show the 
monoidal approach to quantum theory and the generalised topos approach to 
quantum theory have compatible interpretations of states and observables.

In the topos approach to quantum theory 
\cite{Flori2013:Topos,DoeringIsham2008:WhatIsAthing} for a fixed Hilbert space 
$H$,
one takes $\Hilb\Alg(H)$ to be the poset of commutative
$C^*$--subalgebras of $\Hom(H,H)$ considered as a category and
$\Hilb\Algvn(H)$ its subcategory whose objects are the commutative von 
Neumann $C^*$--subalgebras of $\Hom(H,H)$. We will briefly 
discuss a physical interpretation for this definition. Representing physical 
systems by $C^*$--algebras of the form $\Hom(H,H)$ allows us to make 
calculations which accurately predict the outcomes of experiments, however it 
is not at all clear how we are to interpret this algebraic structure, or indeed 
quantum theory in general. According to Bohr's 
interpretation of quantum theory \cite{Bohr1949:DiscussionWithEinstein}, 
although physical reality is by nature quantum, as classical beings conducting 
experiments in our labs we only have 
access to the ``classical parts'' of a quantum system. Much of 
classical physics can be reduced to the study commutative algebras; this 
approach is carefully constructed and motivated in 
\cite{Nestruev2003:SmoothManifoldsAndObservables} where the following picture 
is 
given:

\begin{align*}
\text{Physics lab}&  \qquad\qquad\to&&\text{Commutative unital} \\
& &&\mathbb{R}\text{--algebra } A \\
 \text{Measuring device}& \qquad\qquad\to&&\text{Element of the algebra } A\\
\text{State of the observed}& \qquad\qquad\to&& \text{Homomorphism of unital}  
\\
\text{physical system}& &&\mathbb{R} \text{--algebras } h:A \to \mathbb{R}\\
\text{Output of the}& \qquad\qquad\to&& \text{Value of this function } 
h(a), \\
\text{ measuring device}& && a \in A  \\
\end{align*}
\begin{center}
Figure 1: Algebraic interpretation of classical physics
\end{center}

In \cite{Nestruev2003:SmoothManifoldsAndObservables} the author stresses that 
in 
the interpretation Figure 1. the choice of scalars is unimportant, 
however since
many measurable quantities in classical physics, length, time, energy, etc., 
can be represented by real numbers, 
$\mathbb{R}$ is a reasonable choice. In 
quantum theory the complex numbers are the traditional choice of 
scalars, but one can 
take any ring, or as we will see a semiring in their place.

According to Bohr's interpretation, having access to only the classical 
parts of a quantum system represented by $\Hom(H,H)$ means that we only have 
access to the commutative subalgebras of $\Hom(H,H)$. In the topos approach 
presented in 
\cite{Flori2013:Topos} one considers all of the classical subsystems 
simultaneously by considering the
category of presheaves $[\Hilb\Algvn(H)^{\op}, \,\Set]$, which is a topos. One 
presheaf of 
central importance to the topos approach is characterised by the \emph{Gelfand 
spectrum}. 
Recall the Gelfand spectrum of a commutative $C^*$--algebra $\AlgA$ is the set 
of \emph{characters}
\[\GSpec(\AlgA) = \{ \  \rho:\AlgA \to \mathbb{C} \ | \ \rho \text{ a } 
C^*\text{--algebra homomorphism } \}\]
For a fixed Hilbert space $H$ this defines a functor 
 \[
\begin{tikzpicture}
\node(A) at (0,0) {$\Hilb\Algvn(H)^{\op}$};
\node(B) at (3,0) {$\Set$};
\draw[->](A) to node [above]{$\GSpec$}(B);
\end{tikzpicture}
\]
with the action on morphisms given by restriction. 
By the above physical
interpretation of Figure 1. we see this
functor as assigning to each classical subsystem the set of possible states of 
that system.

While the topos approach introduces new a mathematical language to the study of 
quantum 
theory it still ultimately rests upon the traditional notions of Hilbert spaces 
and von Neumann algebras. The monoidal approach to quantum theory 
\cite{AbramskyCoecke2004:CategoricalSemantics} is an entirely separate approach 
to quantum theory, using different mathematical structures, abstracting away 
from the Hilbert space formalism altogether.

\begin{definition}
A $\dg$\emph{--category} consists of a category $\catA$ together with an 
identity on objects functor 
$\dg: \catA^{\op} 
\to \catA$ satisfying $\dg \circ \dg = \id{\catA}$. A
$\dg$\emph{--symmetric monoidal category} consists of a symmetric monoidal 
category $(\catA,\otimes,I)$ such that: $\catA$ is a $\dg$--category;
$\dg$ is a strict
monoidal functor; and all of the 
symmetric monoidal structure isomorphisms satisfy $\lambda^{-1} = 
\lambda^{\dg}$.
\end{definition}

Symmetric monoidal categories admit a graphical calculus 
\cite{Selinger2011:Survey}, which we assume the reader is familiar with.

The category $\Hilb$ is the archetypal example of a $\dg$--symmetric monoidal 
category. Many aspects of quantum theory can be expressed purely in terms of 
this monoidal structure on the category $\Hilb$. For finite dimensional quantum 
mechanics the notion of an observable can be axiomatised internally by 
commutative 
Frobenius algebras 
\cite{CoeckeEtAl2008:NewDescriptionOrthogonal}, and for infinite 
dimensional quantum mechanics we can encode an observable by a commutative 
$H^*$--algebra \cite{AbramskyHeunen2012:HAlgebras}.
A \emph{concrete} $H^*$\emph{--algebra} 
\cite{Ambrose1945:StructureTheorems} is a (not--necessarily unital) Banach 
algebra such that for each $x \in H$ there is an element $x^* \in H$ such that 
for all $y,z \in H$ 
\[
\langle xy| z \rangle = \langle y | x^* z \rangle
\]

In \cite{AbramskyHeunen2012:HAlgebras} the authors give an axiomatisation for 
$H^*$--algebras in terms of the monoidal structure of the category $\Hilb$, 
which we review in Sect. \ref{sec:InternalAlgebraStructure}. Because they 
generalise Frobenius algebras, $H^*$--algebras are proposed as a 
possibly infinite dimensional notion of ``observable'' in a $\dg$--symmetric 
monoidal category $\catA$.

We are interested in $\dg$--symmetric monoidal categories with 
some additional structure and properties.

\begin{definition}
A $\dg$--symmetric monoidal category $(\catA, \otimes, I)$ is said to be 
\emph{monoidally well--pointed} if for any pair of morphisms $f,g:X \otimes Y 
\to Z$ we have $f\circ 
(x \otimes y) = g \circ (x\otimes y)$ for all $x: I \to X$ and $y:I \to Y$ 
implies $f = g$.
\end{definition}

\begin{definition}
A category $\catA$ is said to have \emph{finite biproducts} if it has a zero 
object $0$, and if for each pair of objects $X_1$ and $X_2$ there exists an 
object 
$X_1 
\oplus X_2$ which is both the coproduct and the product of $X_1$ and $X_2$.
If $\catA$ is a $\dg$--category with finite biproducts such that the 
coprojections $\kappa_i : X_i \to X_1 \oplus X_2$ and projections $\pi_i : 
X_1 \oplus X_2 \to X_i$ are related by $\kappa_i^{\dg} = \pi_i$, then we say 
$\catA$ has \emph{finite} $\dg$\emph{--biproducts}.
\end{definition}

In a category with a zero object $0$, for every pair of objects $X$ and $Y$ 
we call the unique map $X \to 0 \to Y$ the \emph{zero--morphism}, which we 
denote by $0_{X,Y}: X \to Y$, or simply $0:X \to Y$. We say that a pair of 
composable morphisms $f$ and $g$ are \emph{orthogonal} if $f\circ g= 0$.

For a category with finite biproducts each hom-set $\Hom(X,Y)$ is 
equipped with a commutative monoid operation 
\cite[Lemma 18.3]{Mitchell1965:TheoryOfCategories} which we call 
\emph{biproduct convolution}, 
where for
$f,g:X \to Y$ we define $f+g:X \to Y$ by the composition
\begin{equation*}\label{eq:Enrich}
\begin{tikzpicture}[baseline=(current  bounding  box.center)]
\node(A) at (0,0) {$X$};
\node(B) at (1.9,0) {$X\oplus X$};
\node(C) at (4.1,0) {$Y\oplus Y$};
\node(D) at (6,0) {$Y$};
\draw[->](A) to node [above]{$\Delta$}(B);
\draw[->](B) to node [above]{$f \oplus g$}(C);
\draw[->](C) to node [above]{$\nabla$}(D);
\end{tikzpicture}
\end{equation*}
where the additive unit is given by the zero--morphism $0_{X,Y}:X\to Y$.

Categories with finite $\dg$--biproducts admit a matrix 
calculus \cite[Chap. I. Sect. 17.]{Mitchell1965:TheoryOfCategories} 
characterised as follows. For 
$X= \bigoplus\limits_{j=1}^{n} 
X_j$ and $Y= \bigoplus\limits_{i=1}^{m} Y_i$ a morphism $f:X \to Y$ is 
determined 
completely by the morphisms $f_{i,j} : X_i \to Y_j$, and
morphism composition is given by matrix multiplication. If $f$ has matrix 
representation $f_{i,j}$ then $f^\dg$ has matrix representation $f_{j,i}^\dg$.

In Sect. \ref{sec:EndomorphismSemialgebras} we review previous work 
\cite{Dunne2016:NewPerspective} in which we showed how the topos 
approach described above can be generalised away from $\Hilb$ to arbitrary 
$\dg$--symmetric monoidal categories with finite $\dg$--biproducts, and then in 
Sect. \ref{sec:StructureTheorem} we show that this framework is 
versatile enough to easily incorporate aspects 
of the $\dg$--kernel approach to quantum logic 
of Heunen and Jacobs \cite{HeunenJacobs2011:QuantumLogicInDagger}. Integrating 
the $\dg$--kernel framework allows us to prove a structure theorem for 
$H^*$--algebras in 
monoidal categories $\catA$ which generalises the following structure theorem 
of Ambrose for $H^*$--algebras in $\Hilb$ \cite{Ambrose1945:StructureTheorems}.

\begin{theorem}\label{thm:Ambrose}
A concrete commutative $H^*$--algebra $\mu:H \otimes H \to H$ is 
isomorphic to a Hilbert space direct sum 
$\mu \cong \widehat{\bigoplus\limits_{i}}\mu_i$ of one--dimensional algebras 
$\mu_i : \mathbb{C} \otimes \mathbb{C} \to \mathbb{C}$. 
\end{theorem}

\section{Preliminaries}
\label{sec:EndomorphismSemialgebras}

Here we review a construction introduced in 
\cite{Dunne2016:NewPerspective} which generalises the topos approach of 
\cite{DoeringIsham2008:WhatIsAthing,Flori2013:Topos}. This is done using the 
language of semirings, 
semimodules \cite{Golan1992:TheoryOfSemirings}, and semialgebras.

\begin{definition}\label{def:Semiring}
A \emph{semiring} $(R,\cdot, 1 , + ,0)$ consists of a set $R$ equipped with a 
commutative 
monoid operation, \emph{addition}, $+: R \times R \to R$ with unit 
$0\in R$, and a monoid 
operation, \emph{multiplication}, $\cdot : R \times R \to R$, with unit $1\in 
R$, such that $\cdot$ 
distributes over $+$ and $0\cdot s  = s \cdot 0 = 0$ for all $s\in R$.
A semiring is called \emph{commutative} if $\cdot$ is commutative. A 
\emph{$*$--semiring}, or \emph{involutive semiring} is one equipped 
with an operation $*: R \to R$ which is an involution, a homomorphism for 
$(R,+,0)$, and satisfies $(s \cdot t)^* = t^* \cdot s^*$ and $1^* = 1$.
\end{definition}

\begin{definition}
Let $(R,\cdot ,1, +,0)$ be a commutative semiring, an  
$R$--\emph{semimodule} consists 
of a commutative 
monoid $+_M :M\times M \to M$, with unit $0_M$, together with a \emph{scalar 
multiplication} 
$\bullet: 
R \times M \to M$ such that for all $r,s \in R$ and $m,n \in M$:
\begin{multicols}{2}
\begin{enumerate}
 \item $s \bullet (m +_M n) = s\bullet m +_M s \bullet n$ ;
 \item $(r\cdot s) \bullet m = r\bullet ( s \bullet m)$ ;
 \item $(r+s) \bullet m = (r \bullet m) +_M (s \bullet m) $;
 \item $0 \bullet m = s \bullet 0_M = 0_M$;
 \item $1 \bullet m = m$.
\end{enumerate}
\end{multicols}
\end{definition}

\begin{definition}
For $R$ a commutative semiring, an \emph{$R$--semialgebra} 
$(M,\cdot_M,1_M,+_M,0_M)$ consists of an 
$R$--semimodule $(M,+_M,0_M)$ 
equipped with a monoid operation $\cdot_M:M \times M \to M$, with unit $1_M$, 
such that 
$(M,\cdot_M,1_M,+_M,0_M)$ forms a semiring, and where scalar multiplication 
obeys 
$s 
\bullet (m \cdot_M n) = (s
\bullet m )\cdot_M n  =
 m \cdot_M (s \bullet n)$. An $R$--semialgebra is called 
\emph{commutative} if $\cdot_M$ is commutative.
\end{definition}

\begin{definition}\label{def:DaggerSemialgebra}
Let $R$ be a $*$--semiring. An \emph{$R^*$--semialgebra}
consists of an $R$--semialgebra $(M,\cdot_M,1_M,+_M,$ $0_M)$, such that the 
semiring $(M,\cdot_M,1_M,+_M,0_M)$ and $R$ have a compatible involution, i.e. 
one that satisfies $(s \bullet m)^{*}=s^{*} \bullet m^{*}$ for each $s \in R$ 
and $m \in M$.
\end{definition}

Clearly every $*$--semiring $R$ is an $R^*$--semialgebra with scalar 
multiplication taken to be the usual multiplication in $R$. Homomorphisms for 
$R^*$--semialgebras are 
defined in the obvious way. A \emph{unital $R$--subsemialgebra} $i: N 
\hookrightarrow M$ of $M$ 
is a subset $N$ containing $0_M$ and $1_M$ which is closed under all algebraic 
operations. A \emph{subsemialgebra} $N\subset M$ consists of a subset $N$ 
containing $0_M$ and which is closed 
under all algebraic operations making $N$ an $R$--semialgebra in its own right, 
but possibly 
with a different multiplicative unit to $M$. A 
\emph{(unital) $R^*$--subsemialgebra} of a $R^*$--semialgebra is a 
(unital) $R$--subsemialgebra closed 
under taking involutions.

Elements $x$ and $y$ of a semialgebra are said to be 
\emph{orthogonal} if $x\cdot y = 0$. An element $p$ of a
semialgebra is called a \emph{subunital 
idempotent} if $p = p\cdot p$ and there exists $q = q \cdot q$, orthogonal to 
$p$ such that $p + 
q = 1$. A 
\emph{primitive subunital idempotent} $p$ is one such that there are no 
orthogonal
subunital idempotents $s$ and $t$ such that $p = s+t$.  Subunital idempotents 
are also called \emph{weak projections} 
\cite{Harding2008:OrthomodularityInDaggerBiproduct}. 

\begin{theorem}
For a locally small $\dg$--symmetric monoidal category $(\catA, \otimes, I)$ 
with finite $\dg$--biproducts
the 
set $S=\Hom(I,I)$ is a commutative $*$--semiring.
\end{theorem}

The biproduct convolution gives us the additive operation, morphism 
composition gives us the multiplicative operation, and the functor $\dg$ 
provides the involution. It is shown in 
\cite{KellyLaplaza1980:CoherenceForCompact} that this multiplicative operation 
is commutative.

\begin{theorem}
For a locally small $\dg$--symmetric monoidal category $(\catA, \otimes, I)$ 
with finite $\dg$--biproducts where $S= \Hom(I,I)$,
for 
any pair of objects the set $\Hom(X,Y)$ is an $S$--semimodule, and the set 
$\Hom(X,X)$ is a $S^*$--semialgebra.
\end{theorem}

Addition on the set $\Hom(X,Y)$ is given by biproduct convolution. For a 
morphism $f:X \to Y$ the scalar multiplication $s \bullet f$ for $s:I \to I$ is 
defined \cite{Heunen2008:SemimoduleEnrichment} by
\[
\begin{tikzpicture}
\node(A) at (0,0) {$X$};
\node(B) at (1.8,0) {$X\otimes I$};
\node(C) at (4.2,0) {$Y \otimes I$};
\node(D) at (6,0) {$Y$};
\draw[->] (A) to node [above]{$\sim$}(B);
\draw[->] (B) to node [above]{$f \otimes s$}(C);
\draw[->] (C) to node [above]{$\sim$}(D);
\end{tikzpicture}
\]

For $\Hom(X,X)$ multiplication is given by morphism composition and $\dg$ 
provides the involution.

\begin{definition}
For $(\catA, \otimes, I)$ a locally small $\dg$--symmetric monoidal category 
with finite $\dg$--biproducts 
and 
$X$ an object, we define the category $\catA\Alg(X)$ to be the category with 
objects commutative unital $S^*$--subsemialgebras of $\Hom(X,X)$, and arrows 
inclusion 
of unital subsemialgebras.
\end{definition}

For any subset of $B\subset \Hom(X,X)$ the set
$B' = \{\ f:X \to X \ | \ f\circ g= g \circ f \text{ for all } g\in B \ \}$ is 
called the \emph{commutant} of $B$. We define its full subcategory of 
commutative \emph{von Neumann 
$S^*$--subsemialgebras}
\[
\begin{tikzpicture}
\node(A) at (0,0) {$\catA\Algvn(X)$};
\node(B) at (3,0) {$\catA\Alg(X)$};
\draw[right hook->](A) to node [above]{}(B);
\end{tikzpicture}
\]
to have objects those $S^*$--subsemialgebras $\AlgA$ which satisfy the double 
commutant identity $\AlgA = 
\AlgA''$.

The following lemma states some well--known properties of the commutant.
\begin{lemma}\label{lem:CommutantResults}
Let $B$ and $A$ be subsets of $\Hom(X,X)$
\begin{enumerate}
 \item $B'$ is a unital subsemialgebra of $\Hom(X,X)$;
 \item if $B$ is closed under $\dg$ then so is $B'$;
 \item if $A \subset B$ then $B'\subset A'$;
 \item all elements of $B$ commute if and only if $B \subset B'$.
\end{enumerate}
\end{lemma}

An object 
$\AlgA$ of $\catA\Alg(X)$ is \emph{maximal} if it is not properly contained 
in any other commutative subsemialgebra. Being maximal is equivalent to 
satisfying $\AlgA = 
\AlgA'$.

In \cite{Dunne2016:NewPerspective} we gave a direct generalisation of the 
Gelfand 
spectrum of a commutative $C^*$--algebra.
\begin{definition}\label{def:Spectrum}
The \emph{generalised Gelfand spectrum} for the object $X$ in $\catA$ is the 
functor 

\[
\begin{tikzpicture}
\node(A) at (0,0) {$\catA\Alg(X)^{\op}$};
\node(B) at (3,0) {$\Set$};
\draw[->](A) to node [above]{$\GSpec$}(B);
\end{tikzpicture}
\]
defined on objects
$
\GSpec(\AlgA) = \{  \ \rho:\AlgA \to S \ | \ \rho  \text{ an } 
S^*\text{--semialgebra homomorphism } \}$ the set of \emph{characters} on 
$\AlgA$,
while the action on morphism is given by restriction.
\end{definition}

The physical interpretation of Figure 1. remains valid, and we still think of 
the Gelfand spectrum as 
assigning to each classical subsystem the set of possible states of that 
subsystem.
When we take $\catA$ to be the category of Hilbert spaces we obtain exactly the 
category studied in the topos approach to quantum theory 
\cite{DoeringIsham2008:WhatIsAthing,Flori2013:Topos}, and where the 
Gelfand spectrum reduces to the conventional notion.

\begin{remark}\label{rem:KSContextuality}
A principal result of the topos approach 
is that the Kochen--Specker theorem 
\cite{KochenSpecker1975:LogicalStructures} -- which asserts the contextual 
nature 
of quantum theory -- is equivalent to the statement that the Gelfand spectrum 
has 
no global sections \cite[Corollary 9.1]{Flori2013:Topos}. Hence studying the 
global sections of $\GSpec$ allows us to address a more general notion of 
contextuality which we develop in \cite{Dunne2017:SpecPreshKSAndQVR}.
\end{remark}

\begin{remark}
The Gelfand spectrum of a $C^*$--algebra is not just a set, but 
a compact Hausdorff topological space. In \cite{Dunne2017:SpecPreshKSAndQVR} we 
showed that for $\AlgA$ in $\catA\Alg(X)$  the 
Gelfand spectrum $\GSpec(\AlgA)$ comes naturally equipped with the structure of 
a compact topological space.
\end{remark}

\section{From Internal to External Algebraic Structures}
\label{sec:InternalAlgebraStructure}

For a $\dg$--symmetric monoidal category $(\catA, \otimes, I)$, an 
\emph{algebra in} $\catA$ consists of a \emph{carrier} object $X$, and a 
\emph{multiplication} morphism $\mu:X \otimes X \to X$ 
($\begin{tikzpicture}[scale=0.5, 
transform shape]
                                         \node(A) at (0,0) {$%
\beginpgfgraphicnamed{mult}
\InputIfFileExists{mult.tikz}{}{\input{./figures/mult.tikz}}
\endpgfgraphicnamed$};
                                        \end{tikzpicture}$ in the graphical 
calculus). Dually, a 
\emph{coalgebra in} $\catA$ 
consists of a \emph{carrier} object $X$, and a \emph{comultiplication} morphism 
$\delta: X \to X 
\otimes X$ ($\begin{tikzpicture}[scale=0.5, 
transform shape]
                                         \node(A) at (0,0) {$%
\beginpgfgraphicnamed{comult}
\InputIfFileExists{comult.tikz}{}{\input{./figures/comult.tikz}}
\endpgfgraphicnamed$};
                                        \end{tikzpicture}$ in the graphical 
calculus).
An \emph{algebra--coalgebra pair} consists of a carrier object with given 
multiplication and comultiplication maps. Note that in a 
$\dg$--symmetric monoidal category each 
algebra $(X,\mu)$ also defines coalgebra $(X, \mu^\dg)$ and hence every algebra 
in a $\dg$--symmetric monoidal category forms an algebra--coalgebra pair. 
A pair of this type will be referred to as a $\dg$\emph{--algebra}. Consider 
the 
following axioms for a $\dg$--algebra:

\begin{align*}
 \text{ there exists } \quad %
\beginpgfgraphicnamed{unit}
\InputIfFileExists{unit.tikz}{}{\input{./figures/unit.tikz}}
\endpgfgraphicnamed \quad \text{ such 
that } \qquad 
\beginpgfgraphicnamed{unitlaw1}
\InputIfFileExists{unitlaw1.tikz}{}{\input{./figures/unitlaw1.tikz}}
\endpgfgraphicnamed\quad = \quad %
\beginpgfgraphicnamed{unitlaw2}
\InputIfFileExists{unitlaw2.tikz}{}{\input{./figures/unitlaw2.tikz}}
\endpgfgraphicnamed = \quad %
\beginpgfgraphicnamed{id1}
\InputIfFileExists{id1.tikz}{}{\input{./figures/id1.tikz}}
\endpgfgraphicnamed
\qquad \qquad &\text{(U)} \\ \ \qquad \qquad %
\beginpgfgraphicnamed{special1}
\InputIfFileExists{special1.tikz}{}{\input{./figures/special1.tikz}}
\endpgfgraphicnamed \quad = \quad 
\beginpgfgraphicnamed{special2}
\InputIfFileExists{special2.tikz}{}{\input{./figures/special2.tikz}}
\endpgfgraphicnamed 
\qquad \quad &\text{(S)}\\
\beginpgfgraphicnamed{assoc1}
\InputIfFileExists{assoc1.tikz}{}{\input{./figures/assoc1.tikz}}
\endpgfgraphicnamed \quad = \quad %
\beginpgfgraphicnamed{assoc2}
\InputIfFileExists{assoc2.tikz}{}{\input{./figures/assoc2.tikz}}
\endpgfgraphicnamed \qquad &\text{(A)} \\  
\beginpgfgraphicnamed{comm1}
\InputIfFileExists{comm1.tikz}{}{\input{./figures/comm1.tikz}}
\endpgfgraphicnamed \quad = 
\quad %
\beginpgfgraphicnamed{mult}
\InputIfFileExists{mult.tikz}{}{\input{./figures/mult.tikz}}
\endpgfgraphicnamed \quad 
\quad &\text{(C)}\\
\text{for each } \quad %
\beginpgfgraphicnamed{pointx}
\InputIfFileExists{pointx.tikz}{}{\input{./figures/pointx.tikz}}
\endpgfgraphicnamed \quad \text{ there exists } 
\quad %
\beginpgfgraphicnamed{copointxstar}
\InputIfFileExists{copointxstar.tikz}{}{\input{./figures/copointxstar.tikz}}
\endpgfgraphicnamed \quad \text{ such that } \qquad %
\beginpgfgraphicnamed{Haxiom2}
\InputIfFileExists{Haxiom2.tikz}{}{\input{./figures/Haxiom2.tikz}}
\endpgfgraphicnamed 
\quad = \quad %
\beginpgfgraphicnamed{Haxiom1}
\InputIfFileExists{Haxiom1.tikz}{}{\input{./figures/Haxiom1.tikz}}
\endpgfgraphicnamed \qquad \quad &\text{(H)}\\
\end{align*}

%

Note that since we are considering $\dg$--algebras we get the corresponding 
inverted
equations (U), (A) and (C) for the comultiplication morphism.

\begin{definition}
A \emph{commutative $H^*$--algebra} in a $\dg$--symmetric monoidal 
category $\catA$ is a 
$\dg$--algebra satisfying axioms 
(A), (C), (S) and (H). A commutative $H^*$--algebra is said to be \emph{unital} 
if it satisfies (U). 
\end{definition}

It is shown in \cite[Lemma 5.5]{AbramskyHeunen2012:HAlgebras} that in the 
finite--dimensional setting commutative unital $H^*$--algebras serve as a 
axiomatistaion for observables. This is done by showing the close relationship 
between the axiom (H) and the Frobenius axioms, and hence the authors 
propose commutative non--unital $H^*$--algebras as an axiomatisation for 
observables in 
infinite dimensional quantum theory.

For an algebra $(X, \mu)$ the \emph{set--like elements} (or \emph{copyable 
elements}, \emph{group--like elements}, or \emph{classical elements}) are the 
morphisms $\alpha: I \to  X$ satisfying
\[
\beginpgfgraphicnamed{setlike1}
\InputIfFileExists{setlike1.tikz}{}{\input{./figures/setlike1.tikz}}
\endpgfgraphicnamed \qquad = \qquad %
\beginpgfgraphicnamed{setlike2}
\InputIfFileExists{setlike2.tikz}{}{\input{./figures/setlike2.tikz}}
\endpgfgraphicnamed
\]
Under the interpretation of $(X,\mu)$ as an observable, one typically views the 
set--like elements as corresponding with the observable outcomes or states 
associated with that observable. We will require that the set--like 
elements of a $H^*$--algebra satisfy $\alpha^\dg = \tilde{\alpha}$. 
Furthermore we will require that a $H^*$--algebra admits a family of 
\emph{orthonormal} set--like elements, meaning that 
for set--like elements $\alpha$ and $\beta$ the composition $\alpha \circ 
\beta^\dg: I \to I$ is the zero--morphism if $\alpha \not= \beta$ and is the 
identity morphism if $\alpha = \beta$. The cardinality of the set of 
set--like 
elements is the \emph{dimension} of $(X,\mu)$.

The following theorem shows how the notion of observable in the monoidal 
approach to quantum theory -- a commutative $H^*$--algebra $(X,\mu)$ in $\catA$ 
-- lifts naturally to the notion of observable in the generalised topos approach 
-- a commutative von Neumann $S^*$--semialgebra $\AlgX$ in $\catA\Algvn(X)$.

\begin{theorem}\label{thm:GeneratesvonNeumann}
Let $\catA$ be a monoidally well--pointed $\dg$--symmetric monoidal category 
with finite 
$\dg$--biproducts, and let $(X,\mu)$ be an $H^*$--algebra in $\catA$. Consider 
the set of endomorphisms on $X$
\[
R(\mu) = \{ \ \ R_x = %
\beginpgfgraphicnamed{Haxiom2}
\InputIfFileExists{Haxiom2.tikz}{}{\input{./figures/Haxiom2.tikz}}
\endpgfgraphicnamed \quad | \quad \text{for all points } 
\quad  
\beginpgfgraphicnamed{pointx}
\InputIfFileExists{pointx.tikz}{}{\input{./figures/pointx.tikz}}
\endpgfgraphicnamed \quad \}
\]
The commutant $R(\mu)'$ is a 
maximal commutative von Neumann $S^*$--semialgebra. Moreover, if 
$(X,\mu)$ satisfies 
(U) then $R(\mu)=R(\mu)'$.

\proof{ It is easy to verify from axioms (A) and (C) that the 
elements of 
$R(\mu)$ commute with one another, and hence by Lemma 
\ref{lem:CommutantResults}.4. $R(\mu)\subset R(\mu)'$. By the (H) axiom 
$R(\mu)$ is closed under $\dg$ and by Lemma
\ref{lem:CommutantResults}.2. so is $R(\mu)'$. By Lemma
\ref{lem:CommutantResults}.1. $R(\mu)'$ is closed under the algebraic 
operations and hence $R(\mu)'$ 
is a commutative $S^*$--semialgebra. 

The set $R(\mu)'$ is a maximal commutative 
von Neumman $S^*$--semialgebra if and only 
if $R(\mu)' = R(\mu)''$. Since $R(\mu)$ is commutative, Lemma 
\ref{lem:CommutantResults}.4. implies $R(\mu) \subset 
R(\mu)'$, and therefore by Lemma \ref{lem:CommutantResults}.3. $R(\mu)'' 
\subset R(\mu)'$,  and hence to prove maximality of $R(\mu)'$ it is enough to 
show $R(\mu)' \subset 
R(\mu)''$, which by Lemma \ref{lem:CommutantResults}.4. is equivalent to all 
elements of 
$R(\mu)'$ commuting. Consider $h \in R(\mu)'$, then if for 
all $ \, %
\beginpgfgraphicnamed{pointx}
\InputIfFileExists{pointx.tikz}{}{\input{./figures/pointx.tikz}}
\endpgfgraphicnamed$
\begin{equation}\label{eq:LittleResult}
\beginpgfgraphicnamed{vonNeumannAlgebraProof1}
\InputIfFileExists{vonNeumannAlgebraProof1.tikz}{}{\input{./figures/vonNeumannAlgebraProof1.tikz}}
\endpgfgraphicnamed \quad = \quad 
\beginpgfgraphicnamed{vonNeumannAlgebraProof3}
\InputIfFileExists{vonNeumannAlgebraProof3.tikz}{}{\input{./figures/vonNeumannAlgebraProof3.tikz}}
\endpgfgraphicnamed \quad \text{ then by well--pointedness } \quad
\beginpgfgraphicnamed{vonNeumannAlgebraProof2}
\InputIfFileExists{vonNeumannAlgebraProof2.tikz}{}{\input{./figures/vonNeumannAlgebraProof2.tikz}}
\endpgfgraphicnamed \quad = \quad 
\beginpgfgraphicnamed{vonNeumannAlgebraProof4}
\InputIfFileExists{vonNeumannAlgebraProof4.tikz}{}{\input{./figures/vonNeumannAlgebraProof4.tikz}}
\endpgfgraphicnamed
\end{equation}
Hence for $g$ and $h$ 
in $R(\mu)'$ we have
\[
\beginpgfgraphicnamed{phasescommute1}
\InputIfFileExists{phasescommute1.tikz}{}{\input{./figures/phasescommute1.tikz}}
\endpgfgraphicnamed \quad = \quad %
\beginpgfgraphicnamed{phasescommute2}
\InputIfFileExists{phasescommute2.tikz}{}{\input{./figures/phasescommute2.tikz}}
\endpgfgraphicnamed \quad = \quad 
\beginpgfgraphicnamed{phasescommute3}
\InputIfFileExists{phasescommute3.tikz}{}{\input{./figures/phasescommute3.tikz}}
\endpgfgraphicnamed \quad = \quad %
\beginpgfgraphicnamed{phasescommute4}
\InputIfFileExists{phasescommute4.tikz}{}{\input{./figures/phasescommute4.tikz}}
\endpgfgraphicnamed \quad = \quad 
\beginpgfgraphicnamed{phasescommute5}
\InputIfFileExists{phasescommute5.tikz}{}{\input{./figures/phasescommute5.tikz}}
\endpgfgraphicnamed \quad = \quad %
\beginpgfgraphicnamed{phasescommute6}
\InputIfFileExists{phasescommute6.tikz}{}{\input{./figures/phasescommute6.tikz}}
\endpgfgraphicnamed \quad = \quad 
\beginpgfgraphicnamed{phasescommute7}
\InputIfFileExists{phasescommute7.tikz}{}{\input{./figures/phasescommute7.tikz}}
\endpgfgraphicnamed
\]
and hence $R(\mu)' \subset R(\mu)''$, as required.

If $(X,\mu)$ is unital then for each $h \in R(\mu)'$ we have 
\[
\beginpgfgraphicnamed{unitalonthenose1}
\InputIfFileExists{unitalonthenose1.tikz}{}{\input{./figures/unitalonthenose1.tikz}}
\endpgfgraphicnamed \quad = \quad %
\beginpgfgraphicnamed{unitalonthenose2}
\InputIfFileExists{unitalonthenose2.tikz}{}{\input{./figures/unitalonthenose2.tikz}}
\endpgfgraphicnamed \quad = 
\quad %
\beginpgfgraphicnamed{unitalonthenose3}
\InputIfFileExists{unitalonthenose3.tikz}{}{\input{./figures/unitalonthenose3.tikz}}
\endpgfgraphicnamed
\] 
and hence $h \in R(\mu)$, and therefore $R(\mu) = R(\mu)'$, as 
required. \QED}
\end{theorem}

\begin{definition}
Given an $H^*$--algebra $(X,\mu)$ we say that $R(\mu)'$ is the
$S^*$--semialgebra \emph{generated} by $(X,\mu)$.
\end{definition}

Hence an ``observable'' in the monoidal approach -- an $H^*$--algebra -- gives 
rise to an ``observable'' system in the topos approach. Next we show that the 
notion of states in the former -- set--like elements -- determine states in the 
latter -- elements of the Gelfand spectrum.
%

The next theorem shows how the set--like elements naturally form a subset of 
the spectrum.

\begin{theorem}\label{thm:SetLikeInSpectrum}
Let $(X,\mu)$ be an $H^*$--algebra with orthonormal set--like elements and 
$\AlgX$ the von Neumann semialgebra it 
generates. Each set--like element $\alpha$ of $(X,\mu)$ determines an 
$S^*$--semialgebra homomorphism $\rho_\alpha: \AlgX \to S$
defined as
\[
\beginpgfgraphicnamed{Spectrum1}
\InputIfFileExists{Spectrum1.tikz}{}{\input{./figures/Spectrum1.tikz}}
\endpgfgraphicnamed \qquad \mapsto \qquad %
\beginpgfgraphicnamed{Spectrum2}
\InputIfFileExists{Spectrum2.tikz}{}{\input{./figures/Spectrum2.tikz}}
\endpgfgraphicnamed
\]
\proof{
It is easy to check $\rho_\alpha$ preserves zero, the multiplicative unit, 
that it 
respects the dagger, that it preserves biproduct convolution. 
To see $\rho_\alpha$ preserves multiplication consider
\[
\beginpgfgraphicnamed{SpectrumE4}
\InputIfFileExists{SpectrumE4.tikz}{}{\input{./figures/SpectrumE4.tikz}}
\endpgfgraphicnamed \quad = \qquad %
\beginpgfgraphicnamed{SpectrumA4}
\InputIfFileExists{SpectrumA4.tikz}{}{\input{./figures/SpectrumA4.tikz}}
\endpgfgraphicnamed \qquad = \qquad 
\beginpgfgraphicnamed{SpectrumB4}
\InputIfFileExists{SpectrumB4.tikz}{}{\input{./figures/SpectrumB4.tikz}}
\endpgfgraphicnamed \qquad = \qquad %
\beginpgfgraphicnamed{SpectrumC4}
\InputIfFileExists{SpectrumC4.tikz}{}{\input{./figures/SpectrumC4.tikz}}
\endpgfgraphicnamed \quad 
= \qquad 
\beginpgfgraphicnamed{SpectrumD4}
\InputIfFileExists{SpectrumD4.tikz}{}{\input{./figures/SpectrumD4.tikz}}
\endpgfgraphicnamed
\]
and hence $\rho_\alpha (gf) =\rho_\alpha(g) \rho_\alpha(f)$.
\QED}

\end{theorem}

\section{A Structure Theorem for $H^*$--Algebras}
\label{sec:StructureTheorem}
In this section we incorporate concepts from a categorical approach to quantum 
logic \cite{HeunenJacobs2011:QuantumLogicInDagger} into the framework and show 
that in the 
presence of this additional structure the external representations 
of an $H^*$--algebra can be used to determine the structure of that
$H^*$--algebra. This 
theorem is a generalisation of Theorem \ref{thm:Ambrose}, a
structure theorem for concrete $H^*$--algebras.

\begin{definition}
A $\dg$--category $\catA$ with zero object $0$ is said \emph{have 
$\dg$--kernels} if for every morphism $f:X \to Y$ an equaliser $k:K \to X$ of 
$f$ and the zero map $0:X \to Y$ exists and satisfies $k^\dg\circ k = \id{K}$. 
We call $k:K \to X$ the 
\emph{kernel} of $f$.
\end{definition}

Since $\catA$ is a $\dg$--category if it has $\dg$--kernels then it also has 
$\dg$--cokernels, defined dually as a coequaliser.

\begin{definition}
A morphism $f:X \to Y$ in a $\dg$--kernel category is said to be 
\emph{zero--epi} if 
for $g:Y \to Z$ $g \circ f = 0$ implies $g= 0$. A morphism $f$ is said to be 
\emph{zero--mono} if $f^\dg$ is zero--epi.
\end{definition}

It is shown in \cite[Sect. 4.]{HeunenJacobs2011:QuantumLogicInDagger} that in a 
$\dg$--kernel category the collection of zero--epis and $\dg$--kernels form an 
orthogonal factorisation system, and as a consequence every morphism $f:X \to 
Y$ has a 
factorisation $m \circ 
e$ where $m: Z \to Y$ is a $\dg$--kernel and $e:X \to Z$ is zero--epi 
which is unique up to unique isomorphism.

Note that each coprojection $\kappa_X: X \to X \oplus Y$ is a $\dg$--kernel. We 
will require a level of compatibility between the $\dg$--kernel structure and 
the $\dg$--biproduct specified in the following definition.

\begin{definition}
For $\catA$ a $\dg$--category with $\dg$--kernels and finite 
$\dg$--biproducts, we say that the $\dg$--biproducts are \emph{complemented} if 
for every $\dg$--kernel $k: X \to Y$ then there exists $\overline{X}$ such that 
$Y \cong X \oplus \overline{X}$ with 
$k$ the 
first coprojection.
\end{definition}

Equivalently, $\catA$ is complemented if every $\dg$--kernel is a coprojection 
for a $\dg$--biproduct.
Throughout this section we let $\catA$ be a $\dg$--symmetric monoidal category 
with $\dg$--kernels, and finite complemented $\dg$--biproducts.

\begin{lemma}\label{lem:LittleLemma}
Let $\catA$ be a $\dg$--kernel category. Let $f:X 
\to Y$ be a morphism, if $f^\dg \circ f = 0$ then $f=0$.

\proof{let $k \circ e = f$ be the zero--epi $\dg$--kernel factorisation, then 
we have $f^\dg \circ f = e^\dg \circ k^\dg \circ k \circ e = e^\dg \circ e$.
Since $e$ is zero--epi we have $f^\dg 
\circ f= 0 $ implies $e^\dg=0$ which implies $f=0$, as required.
\QED}
\end{lemma}

We call a morphism $f:X \to X$ \emph{normal} if it commutes with its own 
adjoint, that is, $f^\dg\circ f = f \circ f^\dg$. 
Clearly if a morphism $f:X \to X$ belongs to some $\AlgX$ in $\catA\Algvn(X)$ 
then $f$ must be normal. Normal morphisms in the category $\Hilb$ admit a well 
known spectral decomposition. Here we show that normal morphisms in general 
admit a similar structure.

\begin{lemma}\label{lem:LemmaNormalStructure}
Let $\catA$ be a $\dg$--kernel category with finite complemented 
$\dg$--biproducts.
A normal morphism $f:X \to X$ is of the form $f_1 \oplus 0: K\oplus 
\overline{K}\rightarrow K\oplus 
\overline{K}$ for $f_1$ a zero--epi.

\proof{Suppose $f:X \to X$ is normal and let $f = k \circ e$ be its zero--epi 
$\dg$--kernel factorisation where $k:K \to X$.

By the assumption that $\catA$ 
has complemented finite $\dg$--biproducts we have $X \cong K\oplus 
\overline{K}$. 
Hence we have the matrix representation $f= 
 \bigl(\begin{smallmatrix}
  f_1 & f_2 \\
  f_3 & f_4 
 \end{smallmatrix}\bigr)$. It is easy to see that $f_2=0$ and $f_4=0$ and hence 
we have
$ f= 
 \bigl(\begin{smallmatrix}
  f_1 & 0 \\
  f_3 & 0 
 \end{smallmatrix}\bigr)$ and $
 f^\dg = 
 \bigl(\begin{smallmatrix}
  f_1^\dg & f_3^\dg \\
  0 & 0 
 \end{smallmatrix}\bigr)
$.
Therefore
$ f\circ f^\dg =  \bigr(\begin{smallmatrix}
  f_1 f_1^\dg & f_1 f_3^\dg \\
  f_3f_1^\dg & f_3 f_3^\dg 
 \end{smallmatrix}\bigl)$ and $
f^\dg \circ f  =  \bigr(\begin{smallmatrix}
f_1^\dg f_1 + f_3^\dg f_3 & 0 \\
 0 & 0
 \end{smallmatrix}\bigr)$. 
If $f$ is normal then we have $f_3f_3^\dg= 0$, and hence by Lemma 
\ref{lem:LittleLemma} we have 
$f_3=0$ and therefore $f= 
 \bigl(\begin{smallmatrix}
  f_1 & 0 \\
  0 & 0 
 \end{smallmatrix}\bigr)$ as required.
 
The zero--epi $e:X \to K$ has matrix representation $e= 
 \bigl(\begin{smallmatrix}
  f_1 & 0 \\ 
 \end{smallmatrix}\bigr)$ and it is easy to verify that $e$ is zero--epi iff 
$f_1$ is. Hence $f$ has a factorisation $k \circ f_1 \circ k^\dg$, where $k$ is 
a 
$\dg$--kernel, $f_1$ is a zero--epi and $k^\dg$ is a $\dg$--cokernel.
\QED} 
\end{lemma}

A family of morphisms $\{ g_i : X \to Y \}$ is said to be \emph{jointly 
zero--epi} if $f \circ g_i 
= 0$ for all $g_i$, implies $f=0$. We will say that such a jointly zero--epi 
family \emph{forms a cover} of $Y$.

We will ask that the set--like elements of $(X,\mu)$ form a cover for $X$.
The set--like elements of an algebra forming a cover is a far weaker notion 
than that of an algebra having \emph{enough 
set--like elements}, which means that the set--like elements of $(X,\mu)$ 
separate all morphisms out of $X$. For example, $H^*$--algebras in 
$\Rel$ typically do not have enough set--like elements.

In \cite{HeunenJacobs2011:QuantumLogicInDagger} an object $X$ in a 
$\dg$--kernel category is said to be \emph{KSub--simple} if every $\dg$--kernel 
$k:Y \to X$ is an isomorphism. The monoidal units of $\Hilb$ and $\Rel$ are 
KSub--simple. Note that if the monoidal unit of a monoidal category is 
KSub--simple then the semiring of scalars has no zero--divisors.

We now prove an important lemma which lies at the heart of the proof of the 
main structure theorem.

\begin{lemma}\label{thm:indecomposable}
Let $\catA$ be a $\dg$--symmetric monoidal category with $\dg$--kernels such 
that the monoidal unit is KSub--simple. Let $(X,\mu)$ be an $H^*$--algebra in 
$\catA$ with covering orthogonal 
set--like 
elements and let $\AlgX$ be the von Neumann $S^*$--semialgebra generated by 
$(X, \mu)$. There is a set of 
orthogonal 
primitive subunital idempotents in $\AlgX$ corresponding with the set--like 
elements of $(X,\mu)$.

\proof{
Let $\alpha$ be a set--like element and consider $R_\alpha$. By Lemma 
\ref{lem:LemmaNormalStructure} $R_\alpha= 
 \bigl(\begin{smallmatrix}
  \alpha_1 & 0 \\
  0 & 0 
 \end{smallmatrix}\bigr)$, for $\alpha_1$ a zero--epi. Now consider $g$ in $
\AlgX'$, with corresponding matrix 
representation $g = 
 \bigl(\begin{smallmatrix}
  g_1 & g_2 \\
  g_3 & g_4 
 \end{smallmatrix}\bigr)$. Since $g \circ R_\alpha= R_\alpha\circ g$ we 
have $\bigl(\begin{smallmatrix}
  g_1\alpha_1 & 0 \\
  g_3 \alpha_1 & 0 
 \end{smallmatrix}\bigr)
 = \bigl(\begin{smallmatrix}
  \alpha_1 g_1 & \alpha_1 g_2 \\
  0 & 0 
 \end{smallmatrix}\bigr)$. 
Since $\alpha_1$ is zero--epi and $g_3 \alpha = 0$ we conclude $g= 
\bigl(\begin{smallmatrix}
  g_1 & g_2 \\
  0 & g_4 
 \end{smallmatrix}\bigr)$.

 Since $g \circ R_\alpha^\dg= R_\alpha^\dg\circ g$ we 
have $\bigl(\begin{smallmatrix}
  g_1 \alpha_1^\dg & 0 \\
  0 & 0
 \end{smallmatrix}\bigr)
 = \bigl(\begin{smallmatrix}
  \alpha_1^\dg g_1 & \alpha_1^\dg g_2 \\
  0 & 0 
 \end{smallmatrix}\bigr)$. 
and since $\alpha_1^\dg$ is zero--mono we have $g_2=0$ and hence $g= 
\bigl(\begin{smallmatrix}
  g_1 & 0 \\
  0 & g_4 
 \end{smallmatrix}\bigr)$. Clearly for $e_\alpha= 
\bigl(\begin{smallmatrix}
  \id{} & 0 \\
  0 & 0 
 \end{smallmatrix}\bigr)$ and $\overline{e_\alpha}= \bigl(\begin{smallmatrix}
  0 & 0 \\
  0 & \id{} 
 \end{smallmatrix}\bigr)$, we have $g \circ e_\alpha = e_\alpha \circ g= 
\bigl(\begin{smallmatrix}
  g_1 & 0 \\
  0 & 0 
 \end{smallmatrix}\bigr)$ and 
$g 
\circ \overline{e_\alpha} 
=  \overline{e_\alpha} \circ g = 
\bigl(\begin{smallmatrix}
  0 & 0 \\
  0 & g_4 
 \end{smallmatrix}\bigr)$ and hence $e_\alpha$ and $\overline{e_\alpha}$ 
are elements of 
$\AlgX''$ and hence, by Theorem \ref{thm:GeneratesvonNeumann}, are elements of 
$\AlgX$. 
It is easy to verify that the morphisms $e_\alpha$ and $e_\beta$ are orthogonal 
if and only if the set--like elements $\alpha$ and $\beta$ are orthogonal.

We have shown the existence of a family of subunital idempotents $e_\alpha$. It 
remains to show that these subunital idempotents $e_\alpha$ are 
primitive. Suppose there are subunital idempotents $e_1$ and $e_2$ in 
$\AlgX$ such that $e_1 + e_2 = e_\alpha$. Since $e_1$ belongs to $\AlgX$, and 
since 
$\AlgX$ is maximal by Equation \ref{eq:LittleResult} in the proof of Theorem 
\ref{thm:GeneratesvonNeumann}  we have
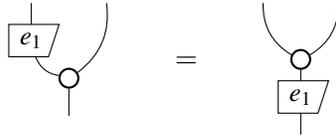
\[
\beginpgfgraphicnamed{NoMoreSetLike1}
\InputIfFileExists{NoMoreSetLike1.tikz}{}{\input{./figures/NoMoreSetLike1.tikz}}
\endpgfgraphicnamed \qquad = \qquad %
\beginpgfgraphicnamed{NoMoreSetLike2}
\InputIfFileExists{NoMoreSetLike2.tikz}{}{\input{./figures/NoMoreSetLike2.tikz}}
\endpgfgraphicnamed
\]
Define $\gamma_1 = e_1 \circ \alpha$ and $\gamma_2 = e_2 \circ \alpha$. It is 
easy to check that $\gamma_1$ and $\gamma_2$ are orthogonal set--like elements 
and that $\gamma_1^\dg = \tilde{\gamma_1}$, and $\gamma_2^\dg = 
\tilde{\gamma_2}$. By Lemma \ref{lem:LittleLemma} $\alpha^\dg \circ \gamma_1$ 
and $\alpha^\dg \circ \gamma_2$ are non--zero, and since $I$ is KSub--simple 
their product is non--zero and therefore
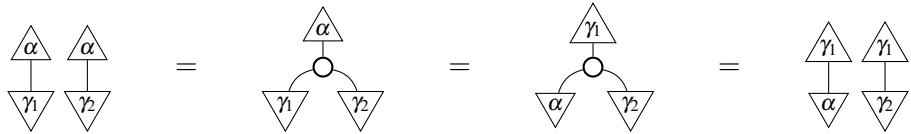
\[
\beginpgfgraphicnamed{NoMoreSetLike1a}
\InputIfFileExists{NoMoreSetLike1a.tikz}{}{\input{./figures/NoMoreSetLike1a.tikz}}
\endpgfgraphicnamed \qquad  = \qquad %
\beginpgfgraphicnamed{NoMoreSetLike2a}
\InputIfFileExists{NoMoreSetLike2a.tikz}{}{\input{./figures/NoMoreSetLike2a.tikz}}
\endpgfgraphicnamed \qquad 
=\qquad %
\beginpgfgraphicnamed{NoMoreSetLike3a}
\InputIfFileExists{NoMoreSetLike3a.tikz}{}{\input{./figures/NoMoreSetLike3a.tikz}}
\endpgfgraphicnamed \qquad  = \qquad 
\beginpgfgraphicnamed{NoMoreSetLike4a}
\InputIfFileExists{NoMoreSetLike4a.tikz}{}{\input{./figures/NoMoreSetLike4a.tikz}}
\endpgfgraphicnamed
\]
but $\gamma_2^\dg \circ \gamma_1$ is zero and hence we have a contradiction, 
and so no such $e_1$ and $e_2$ can exist.
\QED}
\end{lemma}

For $\AlgX$ the $S^*$--semialgebra generated by an $H^*$--algebra $(X,\mu)$, 
Lemma \ref{thm:indecomposable} states that for the elements $e_\alpha$ the 
subsemialgebras $e_i \AlgX \subset \AlgX$ are indecomposable subsemialgebras.

Recall Theorem \ref{thm:Ambrose} states that a $H^*$--algebra is a Hilbert 
space direct sum of one--dimensional algebras. The Hilbert space direct sum is 
not a categorical product or coproduct, but is a \emph{subdirect product} (or 
\emph{subdirect 
union} 
\cite{Birkhoff1944:Subdirect}).
In universal algebra an 
algebraic gadget $A$ 
(e.g. a ring, vector space, semiring, 
lattice or boolean algebra) is a \emph{subdirect product} if there are objects 
$A_i$ such that there is an inclusion $A \hookrightarrow \prod_{i} A_i$ 
such that for each projection $\pi_i: \prod_{i} A_i \to A_i$ the 
composition $A \hookrightarrow \prod_{i} A_i \to A_i$ is surjective. The 
Hilbert space direct sum $\widehat{\bigoplus\limits_i} H_i$ of a 
family of Hilbert spaces $\{H_i\}_i$ has as elements the sequences 
$(x_1,x_2,...)\in \prod_i H_i$ which are square summable, i.e. 
$\sum_i ||x_i||^2 < \infty$, and is a subdirect product of 
vector spaces. The Hilbert space direct sum of a finite family is a finite 
$\dg$--biproduct, however for infinite families of Hilbert spaces the Hilbert 
space direct sum is not a categorical product nor a coproduct, neither of 
which exist for infinite families of Hilbert spaces. The direct sum does retain 
some of the 
properties of a product and coproduct: there are a family of orthogonal 
$\dg$--kernels $\kappa: H_i \to \widehat{\bigoplus\limits_i} H_i$ which are 
jointly epimorphic, that is, if for all $i$ we have $f \circ \pi_i = g \circ 
\pi_i$ 
then 
$f=g$. Given the family $\{ H_i \}_i$ this object 
$\widehat{\bigoplus\limits_i} H_i$ is the unique Hilbert space admitting such a 
family of jointly epimorphic $\dg$--kernels.

\begin{definition}
A $\dg$--kernel category has \emph{sharp $\dg$--kernels} if every jointly 
zero--epi family of $\dg$--kernels is jointly epimorphic. 
\end{definition}

For example, the categories $\Hilb$ and $\Rel$ have sharp $\dg$--kernels.

\begin{definition}
Let $\catA$ be a $\dg$--kernel category. We say 
that 
$\catA$ has \emph{internal direct sums} if for a set of objects $\{X_i\}$ there 
exists an object $\widehat{\bigoplus\limits_i}X_i$  unique up to isomorphism 
together with a family of pairwise 
orthogonal $\dg$--kernels $\kappa_i:X_i \to \widehat{\bigoplus\limits_i} X_i$ 
which are 
jointly epimorphic.

Given families of morphisms $f_i : X_i \to Y_i$, if there exists a morphism 
$\widehat{f}$ such that for each $i$ the diagram
\[
\begin{tikzpicture}
\node(Xi) at (0,0) {$X_i$};
\node(Yi) at (3,0) {$Y_i$};

\node(X) at (0,-1.3) {$\widehat{\bigoplus\limits_i}X_i$};
\node(Y) at (3,-1.3) {$\widehat{\bigoplus\limits_i}Y_i$};

\draw[->](Xi) to node [above]{$f_i$}(Yi);
\draw[dashed, ->](X) to node [above]{$\widehat{f}$}(Y);

\draw[->](Xi) to node [left]{$k_i$}(X);
\draw[->](Yi) to node [right]{$l_i$}(Y);
\end{tikzpicture}
\]
commutes, then we denote $\widehat{f} = \widehat{\bigoplus\limits_i} f_i$.
\end{definition}

In the category $\Hilb$
$\widehat{f}$ exists if and only if there exists $N\in \Nat$ such that $|| f_i 
|| 
\leq N$ 
for all $i$.

It follows from \cite[Proposition 11.6]{AdamekEtAl1990:JoyOfCats} (as an 
internal direct sum is an extremal mono--source) if the 
product 
of the family $\{X_i\}$ exists (e.g. if the family is finite, or if all 
biproducts in $\catA$ exist) then $\widehat{\bigoplus\limits_i}X_i$ and 
$\bigoplus\limits_{i} X_i$ 
coincide.

For the family of indecomposable subsemialgebras $e_i \, \AlgX \subset 
\AlgX$ we can take the coproduct (in the category of $S^*$--semialgebras with 
homomorphisms not necessarily preserving the multiplicative unit) of 
this family $\coproduct_{\, i} \, e_i \, \AlgX$ which consists of sequences of 
elements $(x_1,x_2,...)$ where $x_i \in e_i \AlgX$ such that all but a 
finite number of elements are zero. The product $\prod_{\, i} e_i \AlgX$ 
consists of 
all sequences $(x_1,x_2,...)$ where $x_i \in e_i \AlgX$. We have 
$\coproduct_{\, i} \, e_i \, \AlgX\hookrightarrow \AlgX \hookrightarrow 
\prod_{i} e_i 
\AlgX$, and hence it is easy to see that $\AlgX$ is a subdirect product of its 
indecomposable 
$S^*$--subsemialgebras $e_i \, \AlgX \subset 
\AlgX$.
Clearly if $\catA$ has all biproducts, or if $(X,\mu)$ is finite dimensional 
then $\AlgX \cong \prod_{\, i} \, e_i \, \AlgX$.

Let $(\catA, \otimes, I)$ be a $\dg$--symmetric monoidal category with 
$\dg$-kernels. We say that $\catA$ has \emph{distributive} internal direct sums 
if they satisfy 
$X \otimes (\widehat{\bigoplus\limits_i} Y_i) \cong 
\widehat{\bigoplus\limits_i}(X 
\otimes Y_i)$.

We are now in a position to state and prove the main structure theorem.

\begin{theorem}\label{thm:MainTheoremII}
Let $\catA$ be a $\dg$--symmetric monoidal category with KSub--simple unit, 
sharp $\dg$--kernels 
and 
finite complemented $\dg$--biproducts. Let $\AlgX$ be the $S^*$--semialgebra 
generated by $(X,\mu)$, an $H^*$--algebra in $\catA$ with
covering set--like elements. Let $e_i \AlgX \subset \AlgX$ be the 
indecomposable subsemialgebras on $X$ with $e_i = \id{X_i} \oplus 0$. If 
$\catA$ has distributive internal 
direct sums then $\mu:X \otimes X \to X$ is completely determined by 
an internal direct sum of morphisms $\mu_i:X_i \otimes 
X_i \to X_i$.
\proof{ Let $e_i$ be as in the proof of Lemma \ref{thm:indecomposable}. Each 
$e_i : X \to X$  is of the form $e_i = 
\id{X_i}\oplus 0: X_i \oplus \overline{X_i} \to X_i \oplus \overline{X_i}$ for 
some $\dg$--kernel $k_i :X_i \to X$ and where $e_i = k_i \circ k_i^\dg$. 
Consider the morphisms $\mu \circ (k_i \otimes k_j):X_i \otimes X_j \to X$, we 
have
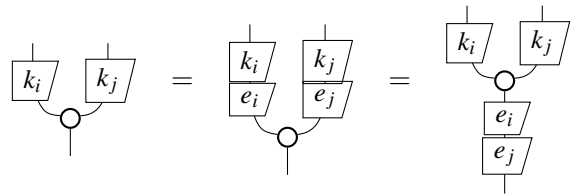
\[
\beginpgfgraphicnamed{InternalProof3}
\InputIfFileExists{InternalProof3.tikz}{}{\input{./figures/InternalProof3.tikz}}
\endpgfgraphicnamed \quad = \quad %
\beginpgfgraphicnamed{InternalProof4}
\InputIfFileExists{InternalProof4.tikz}{}{\input{./figures/InternalProof4.tikz}}
\endpgfgraphicnamed \quad = \quad 
\beginpgfgraphicnamed{InternalProof5}
\InputIfFileExists{InternalProof5.tikz}{}{\input{./figures/InternalProof5.tikz}}
\endpgfgraphicnamed 
\]
and hence if $i\not=j$ then $\mu \circ (k_i \otimes k_j):X_i \otimes X_j \to X$ 
is the zero--morphism.

Now define the 
family of objects $X_{i,j}$, where $X_{i,j} =    X_i$ if  $i=j$ and $X_{i,j} =  
  0$ if $i\not=j$.
  We have a jointly zero--epi family of pairwise 
orthogonal $\dg$--kernels $k_i: X_i \to X$ and hence $X \cong 
\widehat{\bigoplus\limits_{i,j}} X_{i,j}$ and 
if 
internal direct sums are 
distributive we have $X \otimes X \cong 
(\widehat{\bigoplus\limits_i} X_i) \otimes (\widehat{\bigoplus\limits_j} X_j) 
\cong 
\widehat{\bigoplus\limits_{i,j}} (X_i \otimes X_j)$.

Now define $\mu_{i,j}: X_i \otimes X_j \to X_{i,j}$ as $k_i^\dg \circ \mu 
\circ(k_i \otimes k_i)$ if $i=j$ and the zero morphism when $i\not=j$. Then 
$\mu:X \otimes X \to X$ is isomorphic to 
$\widehat{\bigoplus\limits_{i,j}} \mu_{i,j}:\widehat{\bigoplus\limits_{i,j}} 
(X_i \otimes X_j) \to \widehat{\bigoplus\limits_{i,j}} X_{i,j}$. Since the only 
non--zero terms are for $i=j$ the morphism $\mu$ is completely determined 
by the family $\mu_{i}= \mu_{i,i}:X_i \otimes X_i \to X_i$ as claimed.
\QED}
\end{theorem}

\begin{example}
In \cite{AbramskyHeunen2012:HAlgebras} it is shown that a commutative 
$H^*$--algebra $\mu:A \times A \to A$ in $\Rel$, the category of sets and 
relations, is a disjoint union of abelian groups. Applying Theorem 
\ref{thm:MainTheoremII} to $(A,\mu)$, the components $\mu_i:A_i \times A_i \to 
A_i$ are exactly the abelian groups making up $(A,\mu)$.
\end{example}

\bibliography{all}

\end{document}

%% file: preamble.tex
\usepackage[T1]{fontenc}
\usepackage{fixltx2e}
\usepackage{microtype} 
\usepackage{xspace}
\usepackage{multicol}
\usepackage{eurosym}

\makeatletter
\newcommand\etc{etc\@ifnextchar.{}{.\@}\xspace}

\makeatother

\usepackage{bbm}
\usepackage{graphicx}

\newcommand{\inlinegraphic}[2]{
  \dimendef\grafheight=255\dimendef\grafvshift=254
  \grafheight=#1
  \grafvshift=-0.5\grafheight
  \advance\grafvshift by 0.5ex
  \raisebox{\grafvshift}{\includegraphics[height=\grafheight]{images/#2}\xspace}
}

\newcommand{\ninlinegraphic}[2][1.0]{
  \dimendef\grafheight=255\dimendef\grafvshift=254
  \setbox0 = \hbox{\scalebox{#1}{\includegraphics{images/#2}}}
  \grafheight=\the\ht0
  \grafvshift=-0.5\grafheight
  \advance\grafvshift by 0.5ex
  \raisebox{\grafvshift}{\includegraphics[height=\grafheight]{images/#2}\xspace}
}

\newcommand{\QED}{\ensuremath{\null\hfill\square}}

\newcommand{\Alg}{\ensuremath{\textnormal{-\textbf{Alg}}}}
\newcommand{\Algvn}{\ensuremath{\Alg_{\textnormal{vN}}}}

\newcommand{\Spec}{\ensuremath{\textnormal{\text{Spec}}}}

\newcommand{\coproduct}{\mathbin{\rotatebox[origin=c]{180}{$\prod$}}}

\newcommand{\GSpec}{\ensuremath{\Spec_\textnormal{\text{G}}}}

\newcommand{\Hom}{\ensuremath{\textnormal{\text{Hom}}}}

\newcommand{\AlgA}{\ensuremath{\textnormal{\textbf{A}}}}

\newcommand{\AlgX}{\ensuremath{\textnormal{\textbf{X}}}}

\newcommand{\Nat}{\ensuremath{\mathbb{N}}}

\newcommand{\Set}{\ensuremath{\textnormal{\textbf{Set}}}}

\newcommand{\Rel}{\ensuremath{\textnormal{\textbf{Rel}}}}

\newcommand{\Hilb}{\ensuremath{\textnormal{\textbf{Hilb}}}}

\usepackage{latexsym}
\usepackage{amssymb}
\usepackage{amsmath} 
\usepackage{stmaryrd}
\usepackage{mathtools}

\usepackage{amsthm}
\newtheorem{theorem}{Theorem}[section]

\newtheorem{lemma}[theorem]{Lemma}

\theoremstyle{definition}\newtheorem{example}[theorem]{Example}
\theoremstyle{definition}
\theoremstyle{definition}\newtheorem{definition}[theorem]{Definition}
\theoremstyle{definition}
\theoremstyle{definition}\newtheorem{remark}[theorem]{Remark}
\theoremstyle{definition}

\usepackage{diagrams} 
\newarrow{Equals}{}{=}{}{=}{}

\ifx\numericids\undefined
\newcommand{\id}[1]{\ensuremath{\mathrm{id}_{#1}}}
\else
\newcommand{\id}[1]{\ensuremath{1_{#1}}}
\fi

\newcommand{\catA}{\ensuremath{\mathcal{A}}\xspace}

\newcommand{\op}{\ensuremath{^{\mathrm{op}}}\xspace}

\newcommand{\dg}{\ensuremath{\dag}}

\usepackage{hyperref}

%% file: front-matter.tex
\title{On the Structure of Abstract $H^*$--Algebras}
\author{Kevin Dunne
\institute{University of Strathclyde, Glasgow, Scotland.
\email{kevin.dunne@strath.ac.uk}}}



%% file: abstract.tex

\begin{abstract}
\emph{ Previously we have shown that the topos approach to quantum theory of 
Doering and Isham can be generalised to a class of categories typically studied 
within the monoidal approach to quantum theory of Abramsky and Coecke.
In the monoidal approach to quantum theory $H^*$--algebras 
provide an axiomatisation of states and observables. 
Here we show that $H^*$--algebras naturally correspond with the notions of 
states and
observables in the generalised topos approach to quantum theory.
We then combine these results with the $\dg$--kernel approach 
to quantum logic of Heunen and Jacobs, which we use to prove a structure 
theorem for $H^*$--algebras. This structure theorem is a generalisation of the 
structure theorem 
of Ambrose for $H^*$--algebras the category of Hilbert spaces.}
\end{abstract}

%% file: figures/mult.tikz
\begin{tikzpicture}
	\begin{pgfonlayer}{nodelayer}
		\node [style=vertex] (0) at (0, 0) {};
		\node [style=none] (1) at (0, -0.5) {};
		\node [style=none] (2) at (0.25, 0.5) {};
		\node [style=none] (3) at (-0.25, 0.5) {};
	\end{pgfonlayer}
	\begin{pgfonlayer}{edgelayer}
		\draw (1.center) to (0);
		\draw [style=simple, bend left, looseness=1.00] (2.center) to (0);
		\draw [style=simple, bend right, looseness=1.00] (3.center) to (0);
	\end{pgfonlayer}
\end{tikzpicture}

%% file: figures/comult.tikz
\begin{tikzpicture}
	\begin{pgfonlayer}{nodelayer}
		\node [style=vertex] (0) at (0, 0) {};
		\node [style=none] (1) at (0, 0.5) {};
		\node [style=none] (2) at (-0.25, -0.5) {};
		\node [style=none] (3) at (0.25, -0.5) {};
	\end{pgfonlayer}
	\begin{pgfonlayer}{edgelayer}
		\draw (1.center) to (0);
		\draw [style=simple, bend left, looseness=1.00] (2.center) to (0);
		\draw [style=simple, bend right, looseness=1.00] (3.center) to (0);
	\end{pgfonlayer}
\end{tikzpicture}

%% file: figures/unit.tikz
\begin{tikzpicture}
	\begin{pgfonlayer}{nodelayer}
		\node [style=none] (0) at (0, -0.25) {};
		\node [style=vertex] (1) at (0, 0.25) {};
	\end{pgfonlayer}
	\begin{pgfonlayer}{edgelayer}
		\draw [style=simple] (1) to (0.center);
	\end{pgfonlayer}
\end{tikzpicture}

%% file: figures/unitlaw1.tikz
\begin{tikzpicture}
	\begin{pgfonlayer}{nodelayer}
		\node [style=vertex] (0) at (0, 0) {};
		\node [style=vertex] (1) at (0.25, 0.5) {};
		\node [style=none] (2) at (0, -0.5) {};
		\node [style=none] (3) at (-0.25, 0.5) {};
	\end{pgfonlayer}
	\begin{pgfonlayer}{edgelayer}
		\draw (2.center) to (0);
		\draw [style=simple, bend left, looseness=1.00] (1) to (0);
		\draw [style=simple, bend right, looseness=1.00] (3.center) to (0);
	\end{pgfonlayer}
\end{tikzpicture}

%% file: figures/unitlaw2.tikz
\begin{tikzpicture}
	\begin{pgfonlayer}{nodelayer}
		\node [style=vertex] (0) at (0, 0) {};
		\node [style=vertex] (1) at (-0.25, 0.5) {};
		\node [style=none] (2) at (0, -0.5) {};
		\node [style=none] (3) at (0.25, 0.5) {};
	\end{pgfonlayer}
	\begin{pgfonlayer}{edgelayer}
		\draw (2.center) to (0);
		\draw [style=simple, bend right, looseness=1.00] (1) to (0);
		\draw [style=simple, bend left, looseness=1.00] (3.center) to (0);
	\end{pgfonlayer}
\end{tikzpicture}

%% file: figures/id1.tikz
\begin{tikzpicture}
	\begin{pgfonlayer}{nodelayer}
		\node [style=none] (0) at (0, -0.5) {};
		\node [style=none] (1) at (0, 0.5) {};
	\end{pgfonlayer}
	\begin{pgfonlayer}{edgelayer}
		\draw [style=simple] (1.center) to (0.center);
	\end{pgfonlayer}
\end{tikzpicture}

%% file: figures/special1.tikz
\begin{tikzpicture}
	\begin{pgfonlayer}{nodelayer}
		\node [style=vertex] (0) at (0, -0.25) {};
		\node [style=none] (1) at (0, -0.5) {};
		\node [style=vertex] (2) at (0, 0.5) {};
		\node [style=none] (3) at (0, 0.75) {};
	\end{pgfonlayer}
	\begin{pgfonlayer}{edgelayer}
		\draw [style=simple] (0) to (1.center);
		\draw [style=simple] (2) to (3.center);
		\draw [style=simple, bend right, looseness=1.25] (2) to (0);
		\draw [style=simple, bend right, looseness=1.25] (0) to (2);
	\end{pgfonlayer}
\end{tikzpicture}

%% file: figures/special2.tikz
\begin{tikzpicture}
	\begin{pgfonlayer}{nodelayer}
		\node [style=none] (0) at (0, -0.5) {};
		\node [style=none] (1) at (0, 0.5) {};
	\end{pgfonlayer}
	\begin{pgfonlayer}{edgelayer}
		\draw [style=simple] (1.center) to (0.center);
	\end{pgfonlayer}
\end{tikzpicture}

%% file: figures/assoc1.tikz
\begin{tikzpicture}
	\begin{pgfonlayer}{nodelayer}
		\node [style=vertex] (0) at (-0.25, 0.25) {};
		\node [style=vertex] (1) at (0, -0.25) {};
		\node [style=none] (2) at (-0.5, 0.75) {};
		\node [style=none] (3) at (0, 0.75) {};
		\node [style=none] (4) at (0.5, 0.75) {};
		\node [style=none] (5) at (0, -0.75) {};
	\end{pgfonlayer}
	\begin{pgfonlayer}{edgelayer}
		\draw [style=simple, bend right, looseness=1.00] (2.center) to (0);
		\draw [style=simple, bend left, looseness=1.00] (3.center) to (0);
		\draw [style=simple, bend left, looseness=0.75] (4.center) to (1);
		\draw [style=simple] (1) to (5.center);
		\draw [style=simple, bend right=15, looseness=1.00] (0) to (1);
	\end{pgfonlayer}
\end{tikzpicture}

%% file: figures/assoc2.tikz
\begin{tikzpicture}
	\begin{pgfonlayer}{nodelayer}
		\node [style=vertex] (0) at (0.25, 0.25) {};
		\node [style=vertex] (1) at (0, -0.25) {};
		\node [style=none] (2) at (0, 0.75) {};
		\node [style=none] (3) at (0.5, 0.75) {};
		\node [style=none] (4) at (-0.5, 0.75) {};
		\node [style=none] (5) at (0, -0.75) {};
	\end{pgfonlayer}
	\begin{pgfonlayer}{edgelayer}
		\draw [style=simple, bend right, looseness=1.00] (2.center) to (0);
		\draw [style=simple, bend left, looseness=1.00] (3.center) to (0);
		\draw [style=simple, bend right, looseness=0.75] (4.center) to (1);
		\draw [style=simple] (1) to (5.center);
		\draw [style=simple, bend left=15, looseness=1.00] (0) to (1);
	\end{pgfonlayer}
\end{tikzpicture}

%% file: figures/comm1.tikz
\begin{tikzpicture}
	\begin{pgfonlayer}{nodelayer}
		\node [style=vertex] (0) at (0, -0) {};
		\node [style=none] (1) at (0, -0.5) {};
		\node [style=none] (2) at (0.25, 0.75) {};
		\node [style=none] (3) at (-0.25, 0.75) {};
	\end{pgfonlayer}
	\begin{pgfonlayer}{edgelayer}
		\draw [style=simple] (0) to (1.center);
		\draw [style=simple, in=45, out=-90, looseness=1.50] (3.center) to (0);
		\draw [style=simple, in=135, out=-90, looseness=1.50] (2.center) to (0);
	\end{pgfonlayer}
\end{tikzpicture}

%% file: figures/pointx.tikz
\begin{tikzpicture}
	\begin{pgfonlayer}{nodelayer}
		\node [style=none] (0) at (0, -0.25) {};
		\node [style=point] (1) at (0, 0.25) {$x$};
	\end{pgfonlayer}
	\begin{pgfonlayer}{edgelayer}
		\draw [style=simple] (1) to (0.center);
	\end{pgfonlayer}
\end{tikzpicture}

%% file: figures/copointxstar.tikz
\begin{tikzpicture}
	\begin{pgfonlayer}{nodelayer}
		\node [style=none] (0) at (0, 0.5) {};
		\node [style=copoint] (1) at (0, -0) {$\tilde{x}$};
	\end{pgfonlayer}
	\begin{pgfonlayer}{edgelayer}
		\draw [style=simple] (1) to (0.center);
	\end{pgfonlayer}
\end{tikzpicture}

%% file: figures/Haxiom2.tikz
\begin{tikzpicture}
	\begin{pgfonlayer}{nodelayer}
		\node [style=point] (0) at (-0.5, 0.5) {$x$};
		\node [style=vertex] (1) at (0, -0) {};
		\node [style=none] (2) at (0, -0.5) {};
		\node [style=none] (3) at (0.5, 0.5) {};
	\end{pgfonlayer}
	\begin{pgfonlayer}{edgelayer}
		\draw [style=simple] (1) to (2.center);
		\draw [style=simple, bend right, looseness=1.00] (0) to (1);
		\draw [style=simple, bend left, looseness=1.00] (3.center) to (1);
	\end{pgfonlayer}
\end{tikzpicture}

%% file: figures/Haxiom1.tikz
\begin{tikzpicture}
	\begin{pgfonlayer}{nodelayer}
		\node [style=copoint] (0) at (-0.5, -0.25) {$\tilde{x}$};
		\node [style=vertex] (1) at (0, 0.25) {};
		\node [style=none] (2) at (0, 0.75) {};
		\node [style=none] (3) at (0.5, -0.5) {};
	\end{pgfonlayer}
	\begin{pgfonlayer}{edgelayer}
		\draw [style=simple] (1) to (2.center);
		\draw [style=simple, bend left, looseness=1.00] (0) to (1);
		\draw [style=simple, bend left, looseness=1.00] (1) to (3.center);
	\end{pgfonlayer}
\end{tikzpicture}

%% file: figures/setlike1.tikz
\begin{tikzpicture}
	\begin{pgfonlayer}{nodelayer}
		\node [style=point] (0) at (0, 0.5) {$\alpha$};
		\node [style=vertex] (1) at (0, -0) {};
		\node [style=none] (2) at (0.25, -0.5) {};
		\node [style=none] (3) at (-0.25, -0.5) {};
	\end{pgfonlayer}
	\begin{pgfonlayer}{edgelayer}
		\draw [style=simple, bend left, looseness=1.00] (1) to (2.center);
		\draw [style=simple, bend left, looseness=1.00] (3.center) to (1);
		\draw [style=simple] (1) to (0);
	\end{pgfonlayer}
\end{tikzpicture}

%% file: figures/setlike2.tikz
\begin{tikzpicture}
	\begin{pgfonlayer}{nodelayer}
		\node [style=point] (0) at (-0.5, 0.5) {$\alpha$};
		\node [style=none] (1) at (-0.5, -0) {};
		\node [style=none] (2) at (0.25, -0) {};
		\node [style=point] (3) at (0.25, 0.5) {$\alpha$};
	\end{pgfonlayer}
	\begin{pgfonlayer}{edgelayer}
		\draw [style=simple] (1.center) to (0);
		\draw [style=simple] (2.center) to (3);
	\end{pgfonlayer}
\end{tikzpicture}

%% file: figures/vonNeumannAlgebraProof1.tikz
\begin{tikzpicture}
	\begin{pgfonlayer}{nodelayer}
		\node [style=point] (0) at (-0.5, 0.25) {$x$};
		\node [style=vertex] (1) at (0, -0.25) {};
		\node [style=none] (2) at (0, -0.75) {};
		\node [style=map] (3) at (0.5, 0.25) {$h$};
		\node [style=none] (4) at (0.5, 0.75) {};
	\end{pgfonlayer}
	\begin{pgfonlayer}{edgelayer}
		\draw [style=simple] (1) to (2.center);
		\draw [style=simple, bend right, looseness=1.00] (0) to (1);
		\draw [style=simple] (4.center) to (3);
		\draw [style=simple, bend left, looseness=1.00] (3) to (1);
	\end{pgfonlayer}
\end{tikzpicture}

%% file: figures/vonNeumannAlgebraProof3.tikz
\begin{tikzpicture}
	\begin{pgfonlayer}{nodelayer}
		\node [style=point] (0) at (-0.5, 0.5) {$x$};
		\node [style=vertex] (1) at (0, -0) {};
		\node [style=none] (2) at (0.5, 0.5) {};
		\node [style=map] (3) at (0, -0.5) {$h$};
		\node [style=none] (4) at (0, -1) {};
	\end{pgfonlayer}
	\begin{pgfonlayer}{edgelayer}
		\draw [style=simple, bend right, looseness=1.00] (1) to (2.center);
		\draw [style=simple, bend right, looseness=1.00] (0) to (1);
		\draw [style=simple] (4.center) to (3);
		\draw [style=simple] (3) to (1);
	\end{pgfonlayer}
\end{tikzpicture}

%% file: figures/vonNeumannAlgebraProof2.tikz
\begin{tikzpicture}
	\begin{pgfonlayer}{nodelayer}
		\node [style=vertex] (0) at (0, -0.25) {};
		\node [style=none] (1) at (0, -0.75) {};
		\node [style=map] (2) at (0.5, 0.25) {$h$};
		\node [style=none] (3) at (0.5, 0.75) {};
		\node [style=none] (4) at (-0.5, 0.75) {};
	\end{pgfonlayer}
	\begin{pgfonlayer}{edgelayer}
		\draw [style=simple] (0) to (1.center);
		\draw [style=simple] (3.center) to (2);
		\draw [style=simple, bend left, looseness=1.00] (2) to (0);
		\draw [style=simple, bend right, looseness=1.00] (4.center) to (0);
	\end{pgfonlayer}
\end{tikzpicture}

%% file: figures/vonNeumannAlgebraProof4.tikz
\begin{tikzpicture}
	\begin{pgfonlayer}{nodelayer}
		\node [style=vertex] (0) at (0, 0.25) {};
		\node [style=none] (1) at (0.5, 0.75) {};
		\node [style=map] (2) at (0, -0.25) {$h$};
		\node [style=none] (3) at (0, -0.75) {};
		\node [style=none] (4) at (-0.5, 0.75) {};
	\end{pgfonlayer}
	\begin{pgfonlayer}{edgelayer}
		\draw [style=simple, bend right, looseness=1.00] (0) to (1.center);
		\draw [style=simple] (3.center) to (2);
		\draw [style=simple] (2) to (0);
		\draw [style=simple, bend right, looseness=1.00] (4.center) to (0);
	\end{pgfonlayer}
\end{tikzpicture}

%% file: figures/phasescommute1.tikz
\begin{tikzpicture}
	\begin{pgfonlayer}{nodelayer}
		\node [style=map] (0) at (0, 0.5) {$h$};
		\node [style=map] (1) at (0, -0.25) {$g$};
		\node [style=none] (2) at (0, 1) {};
		\node [style=none] (3) at (0, -0.75) {};
	\end{pgfonlayer}
	\begin{pgfonlayer}{edgelayer}
		\draw [style=simple] (0) to (1);
		\draw [style=simple] (0) to (2.center);
		\draw [style=simple] (1) to (3.center);
	\end{pgfonlayer}
\end{tikzpicture}

%% file: figures/phasescommute2.tikz
\begin{tikzpicture}
	\begin{pgfonlayer}{nodelayer}
		\node [style=vertex] (0) at (0, 1) {};
		\node [style=none] (1) at (0, 1.5) {};
		\node [style=vertex] (2) at (0, -0) {};
		\node [style=map] (3) at (0, -0.5) {$h$};
		\node [style=map] (4) at (0, -1) {$g$};
		\node [style=none] (5) at (0, -1.5) {};
	\end{pgfonlayer}
	\begin{pgfonlayer}{edgelayer}
		\draw [style=simple] (1.center) to (0);
		\draw [style=simple] (4) to (5.center);
		\draw [style=simple, bend left=45, looseness=1.25] (0) to (2);
		\draw [style=simple, bend left=45, looseness=1.25] (2) to (0);
		\draw [style=simple] (2) to (3);
		\draw [style=simple] (3) to (4);
	\end{pgfonlayer}
\end{tikzpicture}

%% file: figures/phasescommute3.tikz
\begin{tikzpicture}
	\begin{pgfonlayer}{nodelayer}
		\node [style=map] (0) at (0.5, 0.25) {$h$};
		\node [style=map] (1) at (0, -1) {$g$};
		\node [style=none] (2) at (0, -1.5) {};
		\node [style=vertex] (3) at (0, 1) {};
		\node [style=none] (4) at (0, 1.5) {};
		\node [style=vertex] (5) at (0, -0.5) {};
	\end{pgfonlayer}
	\begin{pgfonlayer}{edgelayer}
		\draw [style=simple] (1) to (2.center);
		\draw [style=simple] (4.center) to (3);
		\draw [style=simple] (5) to (1);
		\draw [style=simple, bend left, looseness=1.00] (3) to (0);
		\draw [style=simple, bend left, looseness=1.00] (0) to (5);
		\draw [style=simple, bend right=60, looseness=1.00] (3) to (5);
	\end{pgfonlayer}
\end{tikzpicture}

%% file: figures/phasescommute4.tikz
\begin{tikzpicture}
	\begin{pgfonlayer}{nodelayer}
		\node [style=map] (0) at (0.5, -0) {$h$};
		\node [style=none] (1) at (0, -1.25) {};
		\node [style=vertex] (2) at (0, 0.75) {};
		\node [style=none] (3) at (0, 1.25) {};
		\node [style=vertex] (4) at (0, -0.75) {};
		\node [style=map] (5) at (-0.5, -0) {$g$};
	\end{pgfonlayer}
	\begin{pgfonlayer}{edgelayer}
		\draw [style=simple] (3.center) to (2);
		\draw [style=simple, bend left, looseness=1.00] (2) to (0);
		\draw [style=simple, bend left, looseness=1.00] (0) to (4);
		\draw [style=simple, bend right, looseness=1.00] (2) to (5);
		\draw [style=simple, bend right, looseness=1.00] (5) to (4);
		\draw [style=simple] (4) to (1.center);
	\end{pgfonlayer}
\end{tikzpicture}

%% file: figures/phasescommute5.tikz
\begin{tikzpicture}
	\begin{pgfonlayer}{nodelayer}
		\node [style=vertex] (0) at (0, 1.25) {};
		\node [style=none] (1) at (0, 1.75) {};
		\node [style=vertex] (2) at (0, -0.25) {};
		\node [style=map] (3) at (-0.5, 0.5) {$g$};
		\node [style=map] (4) at (0, -1) {$h$};
		\node [style=none] (5) at (0, -1.5) {};
	\end{pgfonlayer}
	\begin{pgfonlayer}{edgelayer}
		\draw [style=simple] (1.center) to (0);
		\draw [style=simple, bend right, looseness=1.00] (0) to (3);
		\draw [style=simple, bend right, looseness=1.00] (3) to (2);
		\draw [style=simple] (2) to (4);
		\draw [style=simple] (4) to (5.center);
		\draw [style=simple, bend left=60, looseness=1.25] (0) to (2);
	\end{pgfonlayer}
\end{tikzpicture}

%% file: figures/phasescommute6.tikz
\begin{tikzpicture}
	\begin{pgfonlayer}{nodelayer}
		\node [style=vertex] (0) at (0, 1) {};
		\node [style=none] (1) at (0, 1.5) {};
		\node [style=vertex] (2) at (0, -0) {};
		\node [style=map] (3) at (0, -0.5) {$g$};
		\node [style=map] (4) at (0, -1) {$h$};
		\node [style=none] (5) at (0, -1.5) {};
	\end{pgfonlayer}
	\begin{pgfonlayer}{edgelayer}
		\draw [style=simple] (1.center) to (0);
		\draw [style=simple] (4) to (5.center);
		\draw [style=simple, bend left=45, looseness=1.25] (0) to (2);
		\draw [style=simple, bend left=45, looseness=1.25] (2) to (0);
		\draw [style=simple] (2) to (3);
		\draw [style=simple] (3) to (4);
	\end{pgfonlayer}
\end{tikzpicture}

%% file: figures/phasescommute7.tikz
\begin{tikzpicture}
	\begin{pgfonlayer}{nodelayer}
		\node [style=map] (0) at (0, 0.5) {$g$};
		\node [style=map] (1) at (0, -0.25) {$h$};
		\node [style=none] (2) at (0, 1) {};
		\node [style=none] (3) at (0, -0.75) {};
	\end{pgfonlayer}
	\begin{pgfonlayer}{edgelayer}
		\draw [style=simple] (0) to (1);
		\draw [style=simple] (0) to (2.center);
		\draw [style=simple] (1) to (3.center);
	\end{pgfonlayer}
\end{tikzpicture}

%% file: figures/unitalonthenose1.tikz
\begin{tikzpicture}
	\begin{pgfonlayer}{nodelayer}
		\node [style=map] (0) at (0, -0) {$h$};
		\node [style=none] (1) at (0, 0.5) {};
		\node [style=none] (2) at (0, -0.5) {};
	\end{pgfonlayer}
	\begin{pgfonlayer}{edgelayer}
		\draw [style=simple] (0) to (1.center);
		\draw [style=simple] (0) to (2.center);
	\end{pgfonlayer}
\end{tikzpicture}

%% file: figures/unitalonthenose2.tikz
\begin{tikzpicture}
	\begin{pgfonlayer}{nodelayer}
		\node [style=map] (0) at (0, -0.25) {$h$};
		\node [style=none] (1) at (0, -0.75) {};
		\node [style=vertex] (2) at (0, 0.25) {};
		\node [style=vertex] (3) at (-0.25, 0.75) {};
		\node [style=none] (4) at (0.25, 0.75) {};
	\end{pgfonlayer}
	\begin{pgfonlayer}{edgelayer}
		\draw [style=simple] (0) to (1.center);
		\draw [style=simple, bend right, looseness=1.00] (3) to (2);
		\draw [style=simple, bend right, looseness=0.75] (2) to (4.center);
		\draw [style=simple] (2) to (0);
	\end{pgfonlayer}
\end{tikzpicture}

%% file: figures/unitalonthenose3.tikz
\begin{tikzpicture}
	\begin{pgfonlayer}{nodelayer}
		\node [style=map] (0) at (-0.5, 0.25) {$h$};
		\node [style=none] (1) at (0, -0.75) {};
		\node [style=vertex] (2) at (0, -0.25) {};
		\node [style=vertex] (3) at (-0.5, 0.75) {};
		\node [style=none] (4) at (0.5, 0.75) {};
	\end{pgfonlayer}
	\begin{pgfonlayer}{edgelayer}
		\draw [style=simple, bend right, looseness=0.75] (2) to (4.center);
		\draw [style=simple] (3) to (0);
		\draw [style=simple, bend right, looseness=1.25] (0) to (2);
		\draw [style=simple] (2) to (1.center);
	\end{pgfonlayer}
\end{tikzpicture}

%% file: figures/Spectrum1.tikz
\begin{tikzpicture}
	\begin{pgfonlayer}{nodelayer}
		\node [style=map] (0) at (0, -0) {$f$};
		\node [style=none] (1) at (0, 0.5) {};
		\node [style=none] (2) at (0, -0.5) {};
	\end{pgfonlayer}
	\begin{pgfonlayer}{edgelayer}
		\draw [style=simple] (1.center) to (0);
		\draw [style=simple] (2.center) to (0);
	\end{pgfonlayer}
\end{tikzpicture}

%% file: figures/Spectrum2.tikz
\begin{tikzpicture}
	\begin{pgfonlayer}{nodelayer}
		\node [style=copoint] (0) at (0, -0.5) {$\alpha$};
		\node [style=point] (1) at (0, 0.5) {$\alpha$};
		\node [style=map] (2) at (0, -0) {$f$};
	\end{pgfonlayer}
	\begin{pgfonlayer}{edgelayer}
		\draw [style=simple] (0) to (2);
		\draw [style=simple] (1) to (2);
	\end{pgfonlayer}
\end{tikzpicture}

%% file: figures/SpectrumE4.tikz
\begin{tikzpicture}
	\begin{pgfonlayer}{nodelayer}
		\node [style=copoint] (0) at (0, -0.75) {$\alpha$};
		\node [style=point] (1) at (0, 1) {$\alpha$};
		\node [style=map] (2) at (0, 0.5) {$f$};
		\node [style=map] (3) at (0, -0.25) {$g$};
	\end{pgfonlayer}
	\begin{pgfonlayer}{edgelayer}
		\draw [style=simple] (1) to (2);
		\draw [style=simple] (2) to (3);
		\draw [style=simple] (3) to (0);
	\end{pgfonlayer}
\end{tikzpicture}

%% file: figures/SpectrumA4.tikz
\begin{tikzpicture}
	\begin{pgfonlayer}{nodelayer}
		\node [style=copoint] (0) at (-0.25, -0.75) {$\alpha$};
		\node [style=point] (1) at (-0.25, 1) {$\alpha$};
		\node [style=map] (2) at (-0.25, 0.5) {$f$};
		\node [style=map] (3) at (-0.25, -0.25) {$g$};
		\node [style=point] (4) at (0.5, 0.5) {$\alpha$};
		\node [style=copoint] (5) at (0.5, -0.25) {$\alpha$};
	\end{pgfonlayer}
	\begin{pgfonlayer}{edgelayer}
		\draw [style=simple] (1) to (2);
		\draw [style=simple] (2) to (3);
		\draw [style=simple] (3) to (0);
		\draw [style=simple] (4) to (5);
	\end{pgfonlayer}
\end{tikzpicture}

%% file: figures/SpectrumB4.tikz
\begin{tikzpicture}
	\begin{pgfonlayer}{nodelayer}
		\node [style=copoint] (0) at (-0.5, -1) {$\alpha$};
		\node [style=map] (1) at (-0.5, 0.25) {$f$};
		\node [style=map] (2) at (-0.5, -0.5) {$g$};
		\node [style=point] (3) at (0, 1.25) {$\alpha$};
		\node [style=copoint] (4) at (0.5, 0.25) {$\alpha$};
		\node [style=vertex] (5) at (0, 0.75) {};
	\end{pgfonlayer}
	\begin{pgfonlayer}{edgelayer}
		\draw [style=simple] (1) to (2);
		\draw [style=simple] (2) to (0);
		\draw [style=simple, bend left, looseness=1.00] (5) to (4);
		\draw [style=simple] (3) to (5);
		\draw [style=simple, bend right, looseness=1.00] (5) to (1);
	\end{pgfonlayer}
\end{tikzpicture}

%% file: figures/SpectrumC4.tikz
\begin{tikzpicture}
	\begin{pgfonlayer}{nodelayer}
		\node [style=copoint] (0) at (-0.5, -0.75) {$\alpha$};
		\node [style=map] (1) at (-0.5, -0.25) {$g$};
		\node [style=copoint] (2) at (0.5, -0.75) {$\alpha$};
		\node [style=point] (3) at (0, 0.75) {$\alpha$};
		\node [style=map] (4) at (0.5, -0.25) {$f$};
		\node [style=vertex] (5) at (0, 0.25) {};
	\end{pgfonlayer}
	\begin{pgfonlayer}{edgelayer}
		\draw [style=simple, bend left, looseness=1.00] (5) to (4);
		\draw [style=simple] (3) to (5);
		\draw [style=simple, bend right, looseness=1.00] (5) to (1);
		\draw [style=simple] (1) to (0);
		\draw [style=simple] (4) to (2);
	\end{pgfonlayer}
\end{tikzpicture}

%% file: figures/SpectrumD4.tikz
\begin{tikzpicture}
	\begin{pgfonlayer}{nodelayer}
		\node [style=copoint] (0) at (-0.5, -0.5) {$\alpha$};
		\node [style=map] (1) at (-0.5, -0) {$g$};
		\node [style=copoint] (2) at (0.5, -0.5) {$\alpha$};
		\node [style=point] (3) at (-0.5, 0.5) {$\alpha$};
		\node [style=map] (4) at (0.5, -0) {$f$};
		\node [style=point] (5) at (0.5, 0.5) {$\alpha$};
	\end{pgfonlayer}
	\begin{pgfonlayer}{edgelayer}
		\draw [style=simple] (1) to (0);
		\draw [style=simple] (4) to (2);
		\draw [style=simple] (3) to (1);
		\draw [style=simple] (5) to (4);
	\end{pgfonlayer}
\end{tikzpicture}

%% file: figures/NoMoreSetLike1.tikz
\begin{tikzpicture}
	\begin{pgfonlayer}{nodelayer}
		\node [style=vertex] (0) at (0, -0.25) {};
		\node [style=none] (1) at (0, -0.75) {};
		\node [style=map] (2) at (-0.5, 0.25) {$e_1$};
		\node [style=none] (3) at (-0.5, 0.75) {};
		\node [style=none] (4) at (0.5, 0.75) {};
	\end{pgfonlayer}
	\begin{pgfonlayer}{edgelayer}
		\draw [style=simple] (0) to (1.center);
		\draw [style=simple] (3.center) to (2);
		\draw [style=simple, bend right, looseness=1.00] (2) to (0);
		\draw [style=simple, bend left, looseness=1.00] (4.center) to (0);
	\end{pgfonlayer}
\end{tikzpicture}

%% file: figures/NoMoreSetLike2.tikz
\begin{tikzpicture}
	\begin{pgfonlayer}{nodelayer}
		\node [style=vertex] (0) at (0, -0) {};
		\node [style=none] (1) at (-0.5, 0.75) {};
		\node [style=map] (2) at (0, -0.5) {$e_1$};
		\node [style=none] (3) at (0, -1) {};
		\node [style=none] (4) at (0.5, 0.75) {};
	\end{pgfonlayer}
	\begin{pgfonlayer}{edgelayer}
		\draw [style=simple, bend left, looseness=1.00] (0) to (1.center);
		\draw [style=simple, bend left, looseness=1.00] (4.center) to (0);
		\draw [style=simple] (0) to (2);
		\draw [style=simple] (2) to (3.center);
	\end{pgfonlayer}
\end{tikzpicture}

%% file: figures/NoMoreSetLike1a.tikz
\begin{tikzpicture}
	\begin{pgfonlayer}{nodelayer}
		\node [style=copoint] (0) at (-0.25, -0.5) {$\gamma_1$};
		\node [style=point] (1) at (-0.25, 0.25) {$\alpha$};
		\node [style=copoint] (2) at (0.5, -0.5) {$\gamma_2$};
		\node [style=point] (3) at (0.5, 0.25) {$\alpha$};
	\end{pgfonlayer}
	\begin{pgfonlayer}{edgelayer}
		\draw [style=simple] (1) to (0);
		\draw [style=simple] (3) to (2);
	\end{pgfonlayer}
\end{tikzpicture}

%% file: figures/NoMoreSetLike2a.tikz
\begin{tikzpicture}
	\begin{pgfonlayer}{nodelayer}
		\node [style=copoint] (0) at (-0.5, -0.5) {$\gamma_1$};
		\node [style=point] (1) at (0, 0.5) {$\alpha$};
		\node [style=copoint] (2) at (0.5, -0.5) {$\gamma_2$};
		\node [style=vertex] (3) at (0, -0) {};
	\end{pgfonlayer}
	\begin{pgfonlayer}{edgelayer}
		\draw [style=simple] (1) to (3);
		\draw [style=simple, bend left, looseness=1.00] (0) to (3);
		\draw [style=simple, bend left, looseness=1.00] (3) to (2);
	\end{pgfonlayer}
\end{tikzpicture}

%% file: figures/NoMoreSetLike3a.tikz
\begin{tikzpicture}
	\begin{pgfonlayer}{nodelayer}
		\node [style=copoint] (0) at (-0.5, -0.5) {$\alpha$};
		\node [style=point] (1) at (0, 0.5) {$\gamma_1$};
		\node [style=copoint] (2) at (0.5, -0.5) {$\gamma_2$};
		\node [style=vertex] (3) at (0, -0) {};
	\end{pgfonlayer}
	\begin{pgfonlayer}{edgelayer}
		\draw [style=simple] (1) to (3);
		\draw [style=simple, bend left, looseness=1.00] (0) to (3);
		\draw [style=simple, bend left, looseness=1.00] (3) to (2);
	\end{pgfonlayer}
\end{tikzpicture}

%% file: figures/NoMoreSetLike4a.tikz
\begin{tikzpicture}
	\begin{pgfonlayer}{nodelayer}
		\node [style=copoint] (0) at (-0.25, -0.5) {$\alpha$};
		\node [style=point] (1) at (-0.25, 0.25) {$\gamma_1$};
		\node [style=copoint] (2) at (0.5, -0.5) {$\gamma_2$};
		\node [style=point] (3) at (0.5, 0.25) {$\gamma_1$};
	\end{pgfonlayer}
	\begin{pgfonlayer}{edgelayer}
		\draw [style=simple] (1) to (0);
		\draw [style=simple] (3) to (2);
	\end{pgfonlayer}
\end{tikzpicture}

%% file: figures/InternalProof3.tikz
\begin{tikzpicture}
	\begin{pgfonlayer}{nodelayer}
		\node [style=vertex] (0) at (0, -0.5) {};
		\node [style=none] (1) at (0, -1) {};
		\node [style=map] (2) at (-0.5, -0) {$k_i$};
		\node [style=map] (3) at (0.5, -0) {$k_j$};
		\node [style=none] (4) at (-0.5, 0.5) {};
		\node [style=none] (5) at (0.5, 0.5) {};
	\end{pgfonlayer}
	\begin{pgfonlayer}{edgelayer}
		\draw [style=simple, bend right, looseness=1.25] (2) to (0);
		\draw [style=simple, bend left, looseness=1.00] (3) to (0);
		\draw [style=simple] (0) to (1.center);
		\draw [style=simple] (4.center) to (2);
		\draw [style=simple] (5.center) to (3);
	\end{pgfonlayer}
\end{tikzpicture}

%% file: figures/InternalProof4.tikz
\begin{tikzpicture}
	\begin{pgfonlayer}{nodelayer}
		\node [style=vertex] (0) at (0, -0.75) {};
		\node [style=none] (1) at (0, -1.25) {};
		\node [style=map] (2) at (-0.5, -0.25) {$e_i$};
		\node [style=map] (3) at (0.5, -0.25) {$e_j$};
		\node [style=map] (4) at (-0.5, 0.25) {$k_i$};
		\node [style=map] (5) at (0.5, 0.25) {$k_j$};
		\node [style=none] (6) at (-0.5, 0.75) {};
		\node [style=none] (7) at (0.5, 0.75) {};
	\end{pgfonlayer}
	\begin{pgfonlayer}{edgelayer}
		\draw [style=simple, bend right, looseness=1.25] (2) to (0);
		\draw [style=simple, bend left, looseness=1.00] (3) to (0);
		\draw [style=simple] (0) to (1.center);
		\draw [style=simple] (6.center) to (4);
		\draw [style=simple] (4) to (2);
		\draw [style=simple] (5) to (3);
		\draw [style=simple] (7.center) to (5);
	\end{pgfonlayer}
\end{tikzpicture}

%% file: figures/InternalProof5.tikz
\begin{tikzpicture}
	\begin{pgfonlayer}{nodelayer}
		\node [style=vertex] (0) at (0, -0) {};
		\node [style=map] (1) at (-0.5, 0.5) {$k_i$};
		\node [style=map] (2) at (0.5, 0.5) {$k_j$};
		\node [style=none] (3) at (-0.5, 1) {};
		\node [style=none] (4) at (0.5, 1) {};
		\node [style=map] (5) at (0, -0.5) {$e_i$};
		\node [style=map] (6) at (0, -1) {$e_j$};
		\node [style=none] (7) at (0, -1.5) {};
	\end{pgfonlayer}
	\begin{pgfonlayer}{edgelayer}
		\draw [style=simple, bend right, looseness=1.25] (1) to (0);
		\draw [style=simple, bend left, looseness=1.00] (2) to (0);
		\draw [style=simple] (3.center) to (1);
		\draw [style=simple] (4.center) to (2);
		\draw [style=simple] (0) to (5);
		\draw [style=simple] (5) to (6);
		\draw [style=simple] (6) to (7.center);
	\end{pgfonlayer}
\end{tikzpicture}

%% file: Submission22.bbl
\begin{thebibliography}{10}
\providecommand{\bibitemdeclare}[2]{}
\providecommand{\surnamestart}{}
\providecommand{\surnameend}{}
\providecommand{\urlprefix}{Available at }
\providecommand{\url}[1]{\texttt{#1}}
\providecommand{\href}[2]{\texttt{#2}}
\providecommand{\urlalt}[2]{\href{#1}{#2}}
\providecommand{\doi}[1]{doi:\urlalt{http://dx.doi.org/#1}{#1}}
\providecommand{\bibinfo}[2]{#2}

\bibitemdeclare{inproceedings}{AbramskyCoecke2004:CategoricalSemantics}
\bibitem{AbramskyCoecke2004:CategoricalSemantics}
\bibinfo{author}{Samson \surnamestart Abramsky\surnameend} \&
  \bibinfo{author}{Bob \surnamestart Coecke\surnameend} (\bibinfo{year}{2004}):
  \emph{\bibinfo{title}{A Categorical Semantics of Quantum Protocols}}.
\newblock In: {\sl \bibinfo{booktitle}{Symposium on Logic in Computer
  Science}}, pp. \bibinfo{pages}{415--425}, \doi{10.1109/LICS.2004.1319636}.

\bibitemdeclare{inbook}{AbramskyHeunen2012:HAlgebras}
\bibitem{AbramskyHeunen2012:HAlgebras}
\bibinfo{author}{Samson \surnamestart Abramsky\surnameend} \&
  \bibinfo{author}{Chris \surnamestart Heunen\surnameend}
  (\bibinfo{year}{2012}): \emph{\bibinfo{title}{H*--algebras and nonunital
  Frobenius algebras: First steps in infinite dimensional categorical quantum
  mechanics}}, pp. \bibinfo{pages}{14--37}.
\newblock \bibinfo{volume}{71}, \bibinfo{publisher}{American Mathematical
  Society}, \doi{10.1090/psapm/071}.

\bibitemdeclare{book}{AdamekEtAl1990:JoyOfCats}
\bibitem{AdamekEtAl1990:JoyOfCats}
\bibinfo{author}{Jiri \surnamestart Adamek\surnameend}, \bibinfo{author}{Horst
  \surnamestart Herrlich\surnameend} \& \bibinfo{author}{George~E.
  \surnamestart Strecker\surnameend} (\bibinfo{year}{2009}):
  \emph{\bibinfo{title}{Abstract and Concrete Categories: the Joy of Cats}}.
\newblock \bibinfo{publisher}{Dover}.

\bibitemdeclare{inproceedings}{Ambrose1945:StructureTheorems}
\bibitem{Ambrose1945:StructureTheorems}
\bibinfo{author}{Warren \surnamestart Ambrose\surnameend}
  (\bibinfo{year}{1945}): \emph{\bibinfo{title}{Structure Theorems for a
  Special Class of Banach Algebras}}.
\newblock In: {\sl \bibinfo{booktitle}{Transactions of the American
  Mathematical Society}}, \bibinfo{volume}{57}, pp. \bibinfo{pages}{364--386},
  \doi{10.1090/S0002-9947-1945-0013235-8}.

\bibitemdeclare{article}{Birkhoff1944:Subdirect}
\bibitem{Birkhoff1944:Subdirect}
\bibinfo{author}{Garrett \surnamestart Birkhoff\surnameend}
  (\bibinfo{year}{1944}): \emph{\bibinfo{title}{Subdirect unions in universal
  algebra}}.
\newblock {\sl \bibinfo{journal}{Bull. Amer. Math. Soc.}}
  \bibinfo{volume}{50}(\bibinfo{number}{10}), pp. \bibinfo{pages}{764--768},
  \doi{10.1090/S0002-9904-1944-08235-9}.

\bibitemdeclare{incollection}{Bohr1949:DiscussionWithEinstein}
\bibitem{Bohr1949:DiscussionWithEinstein}
\bibinfo{author}{Niels \surnamestart Bohr\surnameend} (\bibinfo{year}{1949}):
  \emph{\bibinfo{title}{Discussion with {E}instein on Epistemological Problems
  in Atomic Physics}}.
\newblock In \bibinfo{editor}{Paul~Arthur \surnamestart Schilpp\surnameend},
  editor: {\sl \bibinfo{booktitle}{The Library of Living Philosophers, Volume
  7. Albert {E}instein: Philosopher-Scientist}}, \bibinfo{publisher}{Open
  Court}, \doi{10.1016/S1876-0503(08)70379-7}.

\bibitemdeclare{inproceedings}{CoeckeEtAl2008:NewDescriptionOrthogonal}
\bibitem{CoeckeEtAl2008:NewDescriptionOrthogonal}
\bibinfo{author}{Bob \surnamestart Coecke\surnameend}, \bibinfo{author}{Dusko
  \surnamestart Pavlovic\surnameend} \& \bibinfo{author}{Jamie \surnamestart
  Vicary\surnameend} (\bibinfo{year}{2013}): \emph{\bibinfo{title}{A new
  description of orthogonal bases}}.
\newblock In: {\sl \bibinfo{booktitle}{Mathematical Structures in Computer
  Science}}, \bibinfo{volume}{23}, pp. \bibinfo{pages}{555--567},
  \doi{10.1017/S0960129512000047}.

\bibitemdeclare{book}{Conway2000:ACourseInOperatorTheory}
\bibitem{Conway2000:ACourseInOperatorTheory}
\bibinfo{author}{John~B. \surnamestart Conway\surnameend}
  (\bibinfo{year}{2000}): \emph{\bibinfo{title}{A Course in Operator Theory}}.
\newblock {\sl \bibinfo{series}{Graduate Studies in
  Mathematics}}~\bibinfo{volume}{21}, \bibinfo{publisher}{American Mathematical
  Society}, \doi{10.1090/gsm/021}.

\bibitemdeclare{incollection}{DoeringIsham2008:WhatIsAthing}
\bibitem{DoeringIsham2008:WhatIsAthing}
\bibinfo{author}{Andreas \surnamestart Doering\surnameend} \&
  \bibinfo{author}{Chris \surnamestart Isham\surnameend}
  (\bibinfo{year}{2011}): \emph{\bibinfo{title}{What is a Thing?}}
\newblock In \bibinfo{editor}{Bob \surnamestart Coecke\surnameend}, editor:
  {\sl \bibinfo{booktitle}{New Structures in Physics}},
  chapter~\bibinfo{chapter}{13}, \bibinfo{publisher}{Springer},
  \bibinfo{address}{Heidelberg}, pp. \bibinfo{pages}{753--940},
  \doi{10.1007/978-3-642-12821-9\_13}.

\bibitemdeclare{inbook}{Dunne2016:NewPerspective}
\bibitem{Dunne2016:NewPerspective}
\bibinfo{author}{Kevin \surnamestart Dunne\surnameend} (\bibinfo{year}{2017}):
  \emph{\bibinfo{title}{A New Perspective on Observables in the Category of
  Relations: A Spectral Presheaf for Relations}}, pp.
  \bibinfo{pages}{252--264}.
\newblock \bibinfo{publisher}{Springer International Publishing},
  \doi{10.1007/978-3-319-52289-0\_20}.

\bibitemdeclare{inproceedings}{Dunne2017:SpecPreshKSAndQVR}
\bibitem{Dunne2017:SpecPreshKSAndQVR}
\bibinfo{author}{Kevin \surnamestart Dunne\surnameend} (\bibinfo{year}{2017}):
  \emph{\bibinfo{title}{Spectral Presheaves, {K}ochen--{S}pecker Contextuality,
  and Quantale--Valued Relations}}.
\newblock In: {\sl \bibinfo{booktitle}{Quantum Physics and Logic}}.

\bibitemdeclare{book}{Flori2013:Topos}
\bibitem{Flori2013:Topos}
\bibinfo{author}{Cecilia \surnamestart Flori\surnameend}
  (\bibinfo{year}{2013}): \emph{\bibinfo{title}{A First Course in Topos Quantum
  Theory}}.
\newblock \bibinfo{publisher}{Springer-Verlag Berlin Heidelberg},
  \doi{10.1007/978-3-642-35713-8}.

\bibitemdeclare{book}{Golan1992:TheoryOfSemirings}
\bibitem{Golan1992:TheoryOfSemirings}
\bibinfo{author}{Jonathan~S \surnamestart Golan\surnameend}
  (\bibinfo{year}{1992}): \emph{\bibinfo{title}{The Theory of Semirings with
  Applications in Mathematics and Theoretical Computer Science}}.
\newblock \bibinfo{publisher}{Longman Group UK Ltd.}

\bibitemdeclare{article}{Harding2008:OrthomodularityInDaggerBiproduct}
\bibitem{Harding2008:OrthomodularityInDaggerBiproduct}
\bibinfo{author}{John \surnamestart Harding\surnameend} (\bibinfo{year}{2008}):
  \emph{\bibinfo{title}{Orthomodularity in Dagger Biproduct Categories}}.
\newblock {\sl \bibinfo{journal}{Unpublished Manuscript}}.

\bibitemdeclare{inproceedings}{Heunen2008:SemimoduleEnrichment}
\bibitem{Heunen2008:SemimoduleEnrichment}
\bibinfo{author}{Chris \surnamestart Heunen\surnameend} (\bibinfo{year}{2008}):
  \emph{\bibinfo{title}{Semimodule Enrichment}}.
\newblock In: {\sl \bibinfo{booktitle}{Electr. Notes Theor. Comput. Sci.}},
  \bibinfo{volume}{218}, \doi{10.1016/j.entcs.2008.10.012}.

\bibitemdeclare{article}{HeunenJacobs2011:QuantumLogicInDagger}
\bibitem{HeunenJacobs2011:QuantumLogicInDagger}
\bibinfo{author}{Chris \surnamestart Heunen\surnameend} \&
  \bibinfo{author}{Bart \surnamestart Jacobs\surnameend}
  (\bibinfo{year}{2011}): \emph{\bibinfo{title}{Quantum Logic in Dagger Kernel
  Categories}}.
\newblock {\sl \bibinfo{journal}{Electr. Notes Theor. Comput. Sci.}}
  \bibinfo{volume}{270}(\bibinfo{number}{2}), pp. \bibinfo{pages}{79--103},
  \doi{10.1016/j.entcs.2011.01.024}.

\bibitemdeclare{inproceedings}{IshamButterfield1998:AToposPerspective}
\bibitem{IshamButterfield1998:AToposPerspective}
\bibinfo{author}{Chris \surnamestart Isham\surnameend} \&
  \bibinfo{author}{Jeremy \surnamestart Butterfield\surnameend}
  (\bibinfo{year}{1998}): \emph{\bibinfo{title}{A Topos Perspective on the
  Kochen-Specker Theorem: I. Quantum States as Generalised Valuations}}.
\newblock \urlprefix\url{arXiv:quant-ph/9803055}.

\bibitemdeclare{inproceedings}{KellyLaplaza1980:CoherenceForCompact}
\bibitem{KellyLaplaza1980:CoherenceForCompact}
\bibinfo{author}{Gregory~M. \surnamestart Kelly\surnameend} \&
  \bibinfo{author}{Miguel~L. \surnamestart Laplaza\surnameend}
  (\bibinfo{year}{1980}): \emph{\bibinfo{title}{Coherence for Compact Closed
  Categories}}.
\newblock In: {\sl \bibinfo{booktitle}{Journal of Pure and Applied Algebra}},
  \bibinfo{volume}{19}, pp. \bibinfo{pages}{193--213},
  \doi{10.1016/0022-4049(80)90101-2}.

\bibitemdeclare{inproceedings}{KochenSpecker1975:LogicalStructures}
\bibitem{KochenSpecker1975:LogicalStructures}
\bibinfo{author}{S.~\surnamestart Kochen\surnameend} \& \bibinfo{author}{E.~P.
  \surnamestart Specker\surnameend} (\bibinfo{year}{1975}):
  \emph{\bibinfo{title}{Logical Structures Arising in Quantum Theory}}.
\newblock In: {\sl \bibinfo{booktitle}{The Logico-Algebraic Approach to Quantum
  Mechanics}}, pp. \bibinfo{pages}{263--276},
  \doi{10.1007/978-94-010-1795-4\_15}.

\bibitemdeclare{book}{Mitchell1965:TheoryOfCategories}
\bibitem{Mitchell1965:TheoryOfCategories}
\bibinfo{author}{Barry \surnamestart Mitchell\surnameend}
  (\bibinfo{year}{1965}): \emph{\bibinfo{title}{Theory of Categories}}.
\newblock \bibinfo{publisher}{New York Academic Press}.

\bibitemdeclare{book}{Nestruev2003:SmoothManifoldsAndObservables}
\bibitem{Nestruev2003:SmoothManifoldsAndObservables}
\bibinfo{author}{Jet \surnamestart Nestruev\surnameend} (\bibinfo{year}{2003}):
  \emph{\bibinfo{title}{Smooth Manifolds and Observables}}.
\newblock {\sl \bibinfo{series}{Graduate Texts in Mathematics}}
  \bibinfo{volume}{220}, \bibinfo{publisher}{Springer--Verlag New York, Inc.},
  \doi{10.1007/b98871}.

\bibitemdeclare{incollection}{Selinger2011:Survey}
\bibitem{Selinger2011:Survey}
\bibinfo{author}{Peter \surnamestart Selinger\surnameend}
  (\bibinfo{year}{2011}): \emph{\bibinfo{title}{A Survey of Graphical Languages
  for Monoidal Categories}}.
\newblock In \bibinfo{editor}{Bob \surnamestart Coecke\surnameend}, editor:
  {\sl \bibinfo{booktitle}{New Structures in Physics}},
  chapter~\bibinfo{chapter}{4}, \bibinfo{publisher}{Springer},
  \bibinfo{address}{Heidelberg}, pp. \bibinfo{pages}{289--335},
  \doi{10.1007/978-3-642-12821-9\_4}.

\end{thebibliography}
